\PassOptionsToPackage{table}{xcolor}

\documentclass[sigconf,authorversion,nonacm]{acmart}

\PassOptionsToPackage{table,xcdraw}{xcolor} 
\usepackage{xcolor} 

\usepackage{dirtytalk} 
\usepackage[bottom]{footmisc} 
\usepackage{array, booktabs}
\usepackage{tabularray}
\usepackage{makecell}
\usepackage{graphicx}
\usepackage{multirow}
\usepackage{pifont}
\usepackage{colortbl}
\usepackage{stfloats}

\usepackage{fancybox}
\usepackage{xcolor}
\usepackage{textcomp}
\usepackage{amsmath}

\definecolor{myBlue}{RGB}{0, 122, 255}
\definecolor{myBlack}{RGB}{0, 0, 0}
\definecolor{myRed}{RGB}{190,55,74}

\newcommand{\sayit}[1]{{\textit{\say{#1}}}}

\newcommand{\sayhi}[1]{{\textit{\say{#1}}}}

\usepackage{xspace}

\def\sysname{LandmarkLens\xspace}

\usepackage{caption}

\usepackage[inline]{enumitem}

\usepackage{listings}
\usepackage{xcolor} 

\lstnewenvironment{promptbox}
  {\lstset{
    basicstyle=\ttfamily\small,
    breaklines=true,
    columns=flexible,
    backgroundcolor=\color{gray!10},
    frame=single,
    rulecolor=\color{gray!30},
    framerule=0.5pt,
    framesep=6pt,
    xleftmargin=8pt,
    xrightmargin=8pt,
    aboveskip=6pt,
    belowskip=6pt,
  }}
  {}

\usepackage{array}

\definecolor{rowgray}{HTML}{F2F2F2}

\newcolumntype{L}[1]{>{\raggedright\arraybackslash\hyphenpenalty=10000\exhyphenpenalty=10000}p{#1}}
\newcolumntype{M}[1]{>{\raggedright\arraybackslash\hyphenpenalty=10000\exhyphenpenalty=10000}m{#1}}
\newcolumntype{R}[1]{>{\raggedleft\arraybackslash}m{#1}}

\usepackage{multirow}
\newcommand{\sh}{\cellcolor{rowgray}}

\AtBeginDocument{%
  }

\copyrightyear{2026}
\acmYear{2026}
\setcopyright{cc}
\setcctype{by}
\acmConference[UIST '26]{The 39th Annual ACM Symposium on User Interface Software and Technology}{November 02--05, 2026}{Detroit, MI, USA}
\acmBooktitle{The 39th Annual ACM Symposium on User Interface Software and Technology (UIST '26), November 02--05, 2026, Detroit, MI, USA}
\acmDOI{10.1145/3830398.3830505}
\acmISBN{979-8-4007-2856-3/2026/11}

\begin{document}

\sloppy

\title{LandmarkLens: Predicting and Presenting Effective Landmarks for Mixed-Reality Urban Exploration}


\author{Chu Li}
\orcid{0009-0003-7612-6224}
\affiliation{
\institution{Allen School of Computer Science}
  \institution{University of Washington}
  \city{Seattle}
  \country{USA} 
}
\email{chuchuli@cs.washington.edu}

\author{Yotam Sechayk}
\orcid{0009-0002-5286-0080}
\affiliation{
  \institution{The University of Tokyo}
  \city{Tokyo}
  \country{Japan} 
}
\email{sechayk-yotam@g.ecc.u-tokyo.ac.jp}

\author{Jared Hwang}
\orcid{0009-0000-4769-1521}
\affiliation{%
    \institution{Allen School of Computer Science}
    \institution{University of Washington}
  \city{Seattle}
  \country{USA}
}
\email{jaredhwa@cs.washington.edu}

\author{Jon E. Froehlich}
\orcid{0000-0001-8291-3353}
\affiliation{%
    \institution{Allen School of Computer Science}
    \institution{University of Washington}
  \city{Seattle}
  \country{USA}
}
\email{jonf@cs.washington.edu}

\author{Takeo Igarashi}
\affiliation{
  \institution{The University of Tokyo}
  \city{Tokyo}
  \country{Japan} 
}
\orcid{0000-0002-5495-6441}
\email{takeo@acm.org}

\renewcommand{\shortauthors}{Li et al.}

\begin{abstract}
    People with a poor sense of direction (SOD) struggle to build cognitive maps for effective spatial navigation, and existing navigation tools prioritize efficiency over spatial learning. 
To understand how navigation strategies differ by ability, we conducted a landmark attention study with 20 participants (ten good SOD, ten poor SOD) who navigated across four Tokyo neighborhoods in virtual reality (VR).  
We found systematic group differences in both gaze behavior and the types of landmarks they verbally identify as effective.
Based on these findings, we built LandmarkLens, a mixed-reality (MR) navigation system that uses a vision-language model (VLM) to identify and highlight navigation-relevant landmarks. 
A follow-up study with eight poor-SOD participants showed improved performance in scene recognition, suggesting that guided landmark attention can support landmark-level spatial knowledge acquisition for people with poor SOD, a first step toward broader spatial learning.
\end{abstract}



\begin{teaserfigure}
  \includegraphics[width=\textwidth]{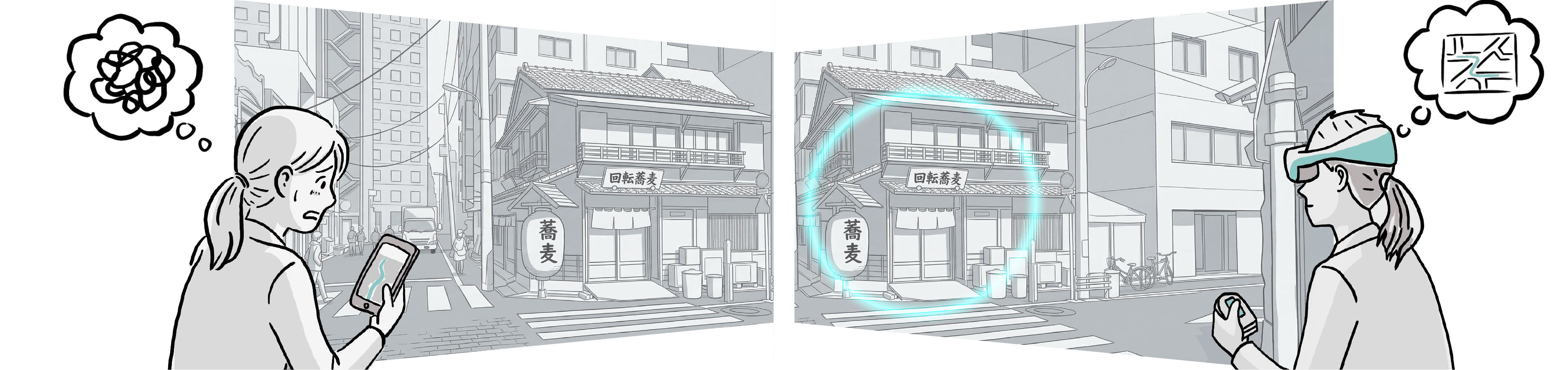}
  \caption{ 
    LandmarkLens highlights navigation-relevant landmarks in a mixed-reality street-view environment. Colored circles mark predicted landmarks; directional arrows guide users toward highlights outside their field of view.
  }
  \Description{A street-view panorama of a Tokyo urban scene as seen through a mixed-reality headset display. Colored neon circles in magenta, orange, and yellow are overlaid on buildings and storefronts to mark predicted navigation landmarks. A directional arrow at the edge of the display points toward a landmark outside the current field of view.
  }
  \vspace{1em}
  \label{fig:teaser}
\end{teaserfigure}


\maketitle

\section{Introduction}

Navigating unfamiliar urban environments is a common and important daily task, yet for people with poor sense of direction (SOD)~\cite{hegarty_development_2002, ishikawa_individual_2023}, it is a source of stress and anxiety~\cite{ishikawa_individual_2023, he_how_2020, topete_how_2024}. 
People with poor SOD are less likely to explore unfamiliar environments~\cite{he_how_2020, ishikawa_individual_2023}, and report greater dependence on GPS-based navigational tools~\cite{topete_how_2024, he_how_2020, hejtmanek_spatial_2018}.
While existing navigation tools guide users from point A to B efficiently, their turn-by-turn directions distract attention away from environmental features~\cite{gardony_how_2013, willis_comparison_2009, hejtmanek_spatial_2018}. 
Over time, this impairs users' spatial knowledge acquisition~\cite{ishikawa_satellite_2019, dahmani_habitual_2020, ruginski_gps_2019}, and undermines the ability to navigate independently~\cite{ishikawa_individual_2023, he_how_2020}.

Prior work has shown that the ability to navigate unfamiliar environments depends on \textit{cognitive mapping}, the process of acquiring, representing, and using knowledge about the surrounding environment~\cite{ishikawa_individual_2023}.
Central to this process are landmarks: objects that are stationary, distinct, and salient, serving as reference points for building mental representations of the environment~\cite{millonig_developing_2007}.
Selecting effective landmarks is correlated with a person's SOD: people with good SOD\footnote{Throughout this paper, we use \sayit{good/poor sense of direction}, \sayit{good/poor navigators}, and \sayit{skilled/unskilled navigators} interchangeably to refer to participants who scored $\geq$5.0 or $\leq$3.5 on the SBSOD scale~\cite{hegarty_development_2002}, respectively.} select fewer but more effective landmarks~\cite{ishikawa_landmark_2012}. 
However, prior work has characterized these differences primarily through verbal self-reports~\cite{ishikawa_landmark_2012}, leaving two open questions: 
(1) how do the \textit{gaze patterns} of good and poor navigators differ during navigation, and 
(2) can guiding poor navigators' attention toward landmarks that skilled navigators rely on improve spatial knowledge acquisition?

To address the first question, we conducted a landmark attention study in virtual reality (VR) with 20 participants, split evenly by SOD as measured by the Santa Barbara Sense-of-Direction (SBSOD) scale~\cite{hegarty_development_2002}. 
Using a custom navigation system rendering $360^\circ$ street-level imagery with integrated eye tracking, participants navigated routes across four Tokyo neighborhoods while verbally describing navigation-relevant features.
We collected concurrent gaze and verbal data, enabling us to analyze not only what participants \textit{said} they noticed but also what they \textit{actually looked at} --- a combination that prior landmark selection studies have not examined.
We found systematic group differences in both gaze behavior and landmark articulation.
Good navigators surveyed a significantly broader horizontal extent of the scene through both wider gaze spread and greater head rotation, and their verbal descriptions disproportionately referenced durable wayfinding cues such as signage and permanent store markers.
Poor navigators, in contrast, more frequently referenced transient and incidental features such as campaign posters, parked bikes, and decorative details.
These findings extend prior verbal-report studies~\cite{ishikawa_landmark_2012} with gaze evidence, confirming that the groups differ not only in \textit{what they articulate} but also in \textit{how they visually scan} the environment.

Building on these findings, we introduce \sysname to address the second question: whether guiding attention toward effective landmarks can improve spatial knowledge acquisition for people with poor SOD.
\sysname is a mixed-reality (MR) navigation system that highlights landmarks informed by the attention patterns and articulation strategies of skilled navigators, layered on top of conventional turn-by-turn directions.
For instance, the system might highlight a neighborhood name sign that people typically overlook, a landmark visible behind you that connects the new street to the one you just left, or a distant tower that orients the pedestrian within the broader cityscape.

To make this approach scalable beyond our study participants, we translated the empirical findings into a landmark prediction pipeline. 
We used gaze distributions from good navigators to crop input imagery to relevant regions, and qualitative insights from their verbal descriptions to construct vision-language model (VLM) prompts that simultaneously promote effective reference points and suppress the transient features poor navigators tend to over-encode.
The resulting MR interface applies the VLM to street-view imagery to predict and present landmarks along a route.
Predicted landmarks are overlaid as interactive highlights.
Directional arrows draw users toward highlights outside their field of view, and a dwell-to-dismiss interaction rewards deliberate attention to each one.
To our knowledge, \sysname is the first system to leverage skilled navigators' gaze behavior and landmark recognition as an attention-guidance mechanism for people with poor SOD.

We evaluated \sysname in a controlled VR study with eight participants who scored $\leq 3.5$ on the SBSOD scale. 
Participants navigated two routes, one using \sysname and one using a turn-by-turn navigation baseline, with route and condition order counterbalanced. Following~\cite{wen_instruction_2014}, we assessed spatial knowledge through three tasks: scene recognition, route choice, and map sketching. 
Participants using \sysname showed significantly higher scene recognition accuracy and recalled nearly twice as many landmarks in map sketches compared to baseline, while route choice accuracy and map structural accuracy remained comparable across conditions.
These results suggest that landmark attention guidance enhances landmark encoding, offering a promising direction for future MR navigation systems that support spatial learning.

In summary, our contributions are:
\begin{enumerate*}
    \item a landmark attention study with 20 participants revealing that good navigators scan a broader horizontal extent of the scene and verbally reference more durable wayfinding cues, while poor navigators disproportionately encode transient and incidental features;
    \item \sysname, an MR navigation system that translates these empirical findings into a VLM-based landmark prediction pipeline and presents navigation-relevant landmarks through gaze-interactive overlays, augmenting turn-by-turn directions to encourage environmental engagement;
    \item a preliminary within-subjects evaluation with eight participants showing that landmark attention guidance significantly improves scene recognition accuracy, with qualitative insights that inform future landmark-based MR navigation;
    \item an open-source dataset of over 3,000 8K $360^\circ$ street-view images with gaze and landmark annotations across Tokyo, supporting future research in HCI, environmental psychology, and spatial cognition.
    All data, code, and study materials are available at \url{https://github.com/makeabilitylab/landmarklens}.
\end{enumerate*}
 \section{Related Work}
We situate our work at the intersection of spatial cognition, individual differences in navigation, and landmark-enhanced MR systems.

\subsection{Landmarks, Routes, and Cognitive Maps}
Our system design and evaluation build on \citet{siegel_development_1975}'s widely adopted trichotomy of spatial knowledge: \textit{landmark knowledge}, recognizing discrete objects or scenes; 
\textit{route knowledge}, encoding landmark sequences and the actions they trigger (\textit{e.g.,} turn left at the church);
and \textit{survey knowledge}, building a \textit{cognitive map} of the environment (\textit{e.g.,}~\cite{wen_individual_2013, wen_instruction_2014, ishikawa_individual_2023, kim_acquisition_2021, montello_new_1998}).

\begin{figure*}[t]
    \centering
    \includegraphics[width=\linewidth]{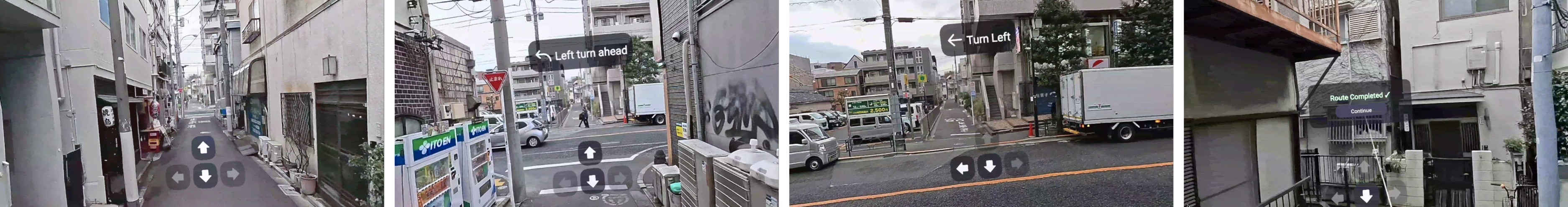}
    \caption{Our custom VR street-view navigation system for the landmark attention study. Directional arrows indicate available movement options, turn notifications appear at intersections. }
    \Description{A screenshot of the VR street-view navigation interface showing an equirectangular panorama of a Tokyo street rendered inside a sphere. Directional arrows on the ground indicate available movement options, and a turn notification overlay is visible near an intersection.}
    \label{fig:attention-study-vr}
\end{figure*}

Central to all three levels are \textit{\textbf{landmarks}}: stationary, distinct, and salient objects that serve as reference points for constructing mental representations of the environment~\cite{millonig_developing_2007}.
Researchers have classified landmarks by saliency --- visual, cognitive, or structural~\cite{sorrows_nature_1999, klippel_structural_2005} --- and by visibility: global landmarks (\textit{e.g.,} mountains or tall buildings) provide broad orientation from afar, while local landmarks aid navigation at close-range decision points~\cite{steck_role_2000}.
\textbf{\textit{Route knowledge}} builds on landmarks through associative memory~\cite{hilton_route_2023}, encoding landmark-place relationships (\textit{when I see the town hall, I am at the town center}~\cite{mallot_place_2018}), place-action pairings (\textit{turn left at the town hall}~\cite{waller_landmarks_2007}), and sequential order (\textit{once I pass the town hall, I will arrive at the restaurant}~\cite{strickrodt_this_2015}).
The most challenging level is \textit{\textbf{survey knowledge}}---a configurational, map-like representation encoding the environment's layout and the spatial relationships among landmarks~\cite{okeefe_hippocampus_1978, tolman_cognitive_1948}.
Acquiring it requires integrating separately learned landmarks and routes within a common frame of reference, demanding understanding that extends beyond any single viewpoint~\cite{wen_instruction_2014, ishikawa_individual_2023, meilinger_integration_2011}.

In \sysname, we evaluate all three levels: following~\citet{wen_instruction_2014}'s task design, our evaluation assesses scene recognition (landmark knowledge), route choice (route knowledge), and map sketching (survey knowledge), enabling us to measure how landmark attention guidance affects spatial learning at each level.

\subsection{Individual Differences in Cognitive Mapping}
Individuals differ widely in cognitive mapping ability~\cite{weisberg_how_2016, wen_individual_2013, weisberg_variations_2014, ishikawa_individual_2023, ishikawa_spatial_2006, hegarty_development_2002}, from expert navigators~\cite{maguire_london_2006} to those with severe navigational impairments~\cite{iaria_developmental_2010}.
The Santa Barbara Sense-of-Direction Scale~\cite{hegarty_development_2002}, a self-report questionnaire assessing navigational abilities on a 7-point scale, has become the most widely used instrument for measuring these differences~\cite{ishikawa_landmark_2012, wen_individual_2013, ishikawa_spatial_2006, wen_instruction_2014, weisberg_variations_2014}.
For reference, a cross-cultural validation ($N = 550$) found a population mean of 4.2 ($SD = 1.1$) on the SBSOD scale~\cite{montello_linguistic_2011}.

These differences have practical consequences: poor cognitive mapping leads to spatial anxiety~\cite{lawton_gender_1994} and over-reliance on GPS-based navigation tools~\cite{he_how_2020}, which in turn draw attention away from environmental features~\cite{gardony_how_2013, gardony_navigational_2015}, reduce active route planning~\cite{willis_comparison_2009}, and impair spatial learning over the long term~\cite{ruginski_gps_2019} --- creating a vicious cycle in which poor navigators become increasingly dependent on tools that further attenuate their spatial learning.

Researchers have explored interventions including repeated route exposure~\cite{uttal_malleability_2013}, practice with feedback~\cite{ishikawa_improving_2020}, and verbalization during learning~\cite{chi_eliciting_1994}.
Of particular relevance is the role of landmark attention: \citet{ishikawa_landmark_2012} found that people with good SOD select fewer but more effective landmarks, while people with poor SOD select unreliable features.
This raises a natural question: can instructing people to attend to effective landmarks facilitate cognitive mapping? To date, this remains untested~\cite{ishikawa_individual_2023}. We designed \sysname to investigate this possibility.

\subsection{Landmark Enhancements in MR}
Landmark-based navigation systems effectively support wayfinding and spatial knowledge acquisition~\cite{wunderlich_landmark-based_2021, goodman_how_2005}, and augmented reality (AR) takes this further by integrating virtual cues into the user's real-world view~\cite{dey_systematic_2018}.
For indoor navigation, researchers have used AR to enhance landmarks for mental map development~\cite{zhang_enhancing_2021}, label semantic landmarks to promote spatial learning~\cite{liu_spatial_2021}, support older adults with landmark-based instructions~\cite{qiu_navmarkar_2024}, and augment landmark perception for people with low vision~\cite{chen_visimark_2025}.
For outdoor urban environments, commercial tools such as Google Maps Live View superimpose only arrows and text, while recent research has explored landmark-based approaches including 3D virtual landmark maps~\cite{zhang_exploring_2022}, destination landmark overlays~\cite{mazurkiewicz_beear_2023}, and detection models trained on researcher-curated salient landmarks~\cite{rao_landmarks_2020}.

A key question across these systems is how landmarks are selected.
In much prior work, selection has been determined by researchers based on general saliency principles (\textit{e.g.,}~\cite{qiu_navmarkar_2024}), while more recent approaches have adapted selection based on user familiarity~\cite{zhu_personalized_2022}, formative studies with target users~\cite{chen_visimark_2025}, or computational detection models~\cite{rao_landmarks_2020}.
However, none of these approaches ground landmark selection in the cognitive and perceptual differences between good and poor navigators.
\sysname bridges this gap by identifying landmark attention patterns from people with good SOD and using them to guide people with poor SOD, directly addressing the open question of whether attention to effective landmarks can facilitate spatial knowledge acquisition~\cite{ishikawa_individual_2023}.
\section{Landmark Attention Study}
\label{sec:landmark-attention-study}

To understand how people with good versus poor SOD attend to landmarks during urban navigation, we conducted a study combining eye gaze tracking and verbal descriptions.
Prior work has examined landmark selection through verbal reports alone~\cite{ishikawa_landmark_2012}; our study extends this by collecting concurrent gaze and verbal data from 20 participants navigating street-view routes across four Tokyo neighborhoods in VR.

\subsection{Participants}
We recruited 20 participants through university mailing lists and snowball sampling.
Potential participants completed a screening survey that included the SBSOD questionnaire~\cite{hegarty_development_2002}.
We selected the first ten respondents who scored $ \geq 5.0$ (good SOD) and the first ten who scored $ \leq 3.5$ (poor SOD); these cutoffs are consistent with~\cite{ishikawa_improving_2020} and fall above and below the population mean reported for the SBSOD scale~\cite{montello_linguistic_2011}.
The mean SBSOD score was 6.0 ($SD = 0.7$) for the good group and 2.9 ($SD = 0.5$) for the poor group.

Participants had a mean age of 26.9 ($SD = 2.6$) in the good group and 28.2 ($SD = 4.1$) in the poor group.
The good group included two women and eight men; the poor group included six women and four men.
Participants were compensated with a JPY \textyen5,000 gift card.
See Appendix \autoref{tab:attention-study-participants} for full participant information.

\subsection{Apparatus}
\textbf{Street-Level Imagery.}
We collected $360^\circ$ street-level imagery from four Tokyo neighborhoods, selected based on~\citet{almazan_emergent_2022} categorization of Tokyo's urban archetypes: Nezu (\textit{Village Tokyo}), Koenji (\textit{Local Tokyo}), Ginza (\textit{Office Tower Tokyo}), and Kinshicho (\textit{Mercantile Tokyo}).
We captured $360^\circ$ timelapse video at 8K resolution using GoPro Max2.
The videos were segmented into equirectangular photo frames at 0.1 second intervals, yielding a total of 3,120 panoramic images.

\noindent\textbf{Route Selection.}
Each neighborhood contained 11 routes: one common route walked by all 20 participants, and ten unique routes assigned one per participant (\autoref{fig:attention-study-routes}).
A total number of 2,377 panoramas were included.
Common routes enable direct gaze comparison across participants, while the unique routes maximize imagery diversity.
Routes were balanced for number of images and number of turns within each neighborhood.
Each participant navigated eight routes (one common and one unique route per neighborhood).
To mitigate ordering effects on gaze data, route order was counterbalanced across participants using Latin square design. 
See Appendix~\autoref{tab:route-order} for the full route presentation order.

\noindent\textbf{VR System.}
We developed a custom street-view navigation application in Unity, deployed on a Samsung Galaxy XR headset (Qualcomm Snapdragon XR2+ Gen 2, $3552 \times 3840$ per-eye micro-OLED display, $109^\circ$ horizontal field of view, four eye-tracking cameras).
The application renders each equirectangular panorama on the interior of a sphere surrounding the user.
Participants navigated using a handheld controller.
Transitions between panoramas were rendered as cross-fade animations, with a turn animation applied when changing direction at intersections.
Directional arrows indicated available movement options at each panorama, and turn notification overlays (\textit{e.g.,} \sayhi{Turn right}, \sayhi{Right turn ahead}) appeared when approaching intersections to guide participants along the route.
The system logged eye gaze origin, gaze direction, gaze hit point on the
panorama (UV coordinates), head position, head rotation, and input events
at the headset's native sampling rate (72\,Hz); 
audio was recorded using an external clip-on microphone.

\noindent\subsection{Procedure}
\textbf{Introduction.}
Following a two-participant pilot that informed our introduction script and task guidance, each session began with an introduction to the study and informed consent.
The researcher showed participants screenshots explaining the navigation controls.
Participants were then fitted with a Samsung Galaxy XR headset and eye gaze tracking was calibrated.
A gaze visualization sphere was briefly shown to confirm tracking accuracy, then turned off.

\noindent\textbf{Practice.}
Participants completed a short practice route (26 images from an unused Nezu route) to familiarize themselves with the system.
They were given the task instructions: \sayit{Imagine you are giving directions to someone who is visiting this area for the first time. As you navigate each route, please verbally describe any features along the way that you think would be helpful as navigation clues.}

\noindent\textbf{Routes.} 
Participants completed eight routes. 
After each route, participants removed the headset, verbally described the route they had just taken including any visual clues and directions they could recall.
As pilot participants preferred sketching during recall, we also provided pen and paper.

\noindent\textbf{Retrace.} 
Participants were informed at the outset that one route would be randomly selected for them to navigate again from memory. 
In practice, this was always the final route, ensuring participants remained attentive to navigation cues throughout all routes.

\noindent\textbf{Debriefing.}
After all routes were completed, we debriefed participants, answered any questions about the study, and showed them the routes on a map upon request. 

\subsection{Analysis}
We analyzed participants' gaze behavior along four dimensions: gaze spread, fixation rate, fixation duration, and head rotation. 
For gaze spread, we computed the standard deviation of gaze hit coordinates on the equirectangular panorama. 
For fixation detection, we used the I-DT algorithm~\cite{salvucci_identifying_2000} with a dispersion threshold of 0.02 in UV space and a minimum duration of 150\,ms. 
We fitted linear mixed models for gaze spread and head rotation, and Gamma generalized linear mixed models for fixation rate and duration.
For the retrace task, we modeled intersection-level accuracy using a binomial generalized linear mixed model with group as a fixed effect and route and participant as random intercepts.
One participant (A17) was identified as an outlier via the $1.5 \times \text{IQR}$ criterion on session duration, gaze spread, and head rotation, and was excluded from all quantitative analyses, yielding a final sample of ten good and nine poor navigators.
 
We transcribed participants' verbal landmark descriptions and analyzed them using a hybrid deductive-inductive coding approach~\cite{charmaz_constructing_2006}. 
One researcher developed an initial codebook based on landmark categories from prior work~\cite{ishikawa_landmark_2012, sorrows_nature_1999}, then iteratively refined it as new codes emerged. 
To establish reliability, a second researcher independently coded two transcripts, one from each SOD group. 
Disagreements were discussed until consensus was reached.
The remaining 18 transcripts were then recoded using the revised codebook.

\begin{figure*}[t]
    \centering
    \includegraphics[width=\linewidth]{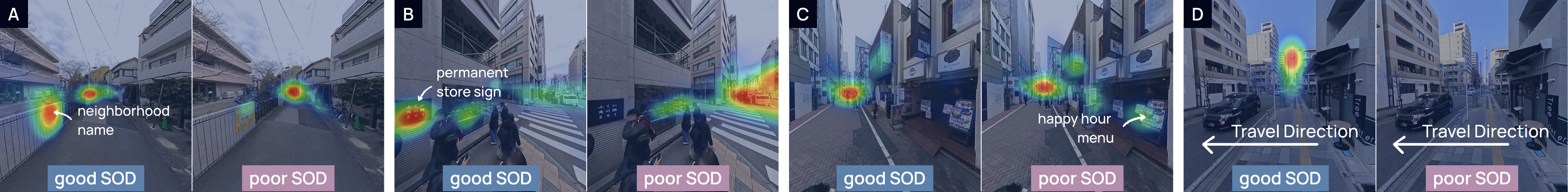}
    \caption{Examples of divergent gaze patterns. 
    Good navigators attended to (A) wayfinding signage and (B) permanent store identifiers. Poor navigators fixated on (C) menus and temporary signage. (D) Good navigators turned to fixate on the Tokyo Skytree as a global landmark; poor navigators did not.
    }
    \Description{Four side-by-side panorama comparisons labeled A through D showing divergent gaze patterns between good and poor navigators. (A) shows gaze concentrated on a wayfinding sign, (B) on a permanent store name engraved in stone, (C) on temporary menus and store signage, and (D) on the Tokyo Skytree where good navigators turned their heads to look at the tower while poor navigators did not.}
    \label{fig:gaze-heatmaps-comparison}
    \vspace{-1em}
\end{figure*}

\begin{figure}[b]
    \centering
    \includegraphics[width=\linewidth]{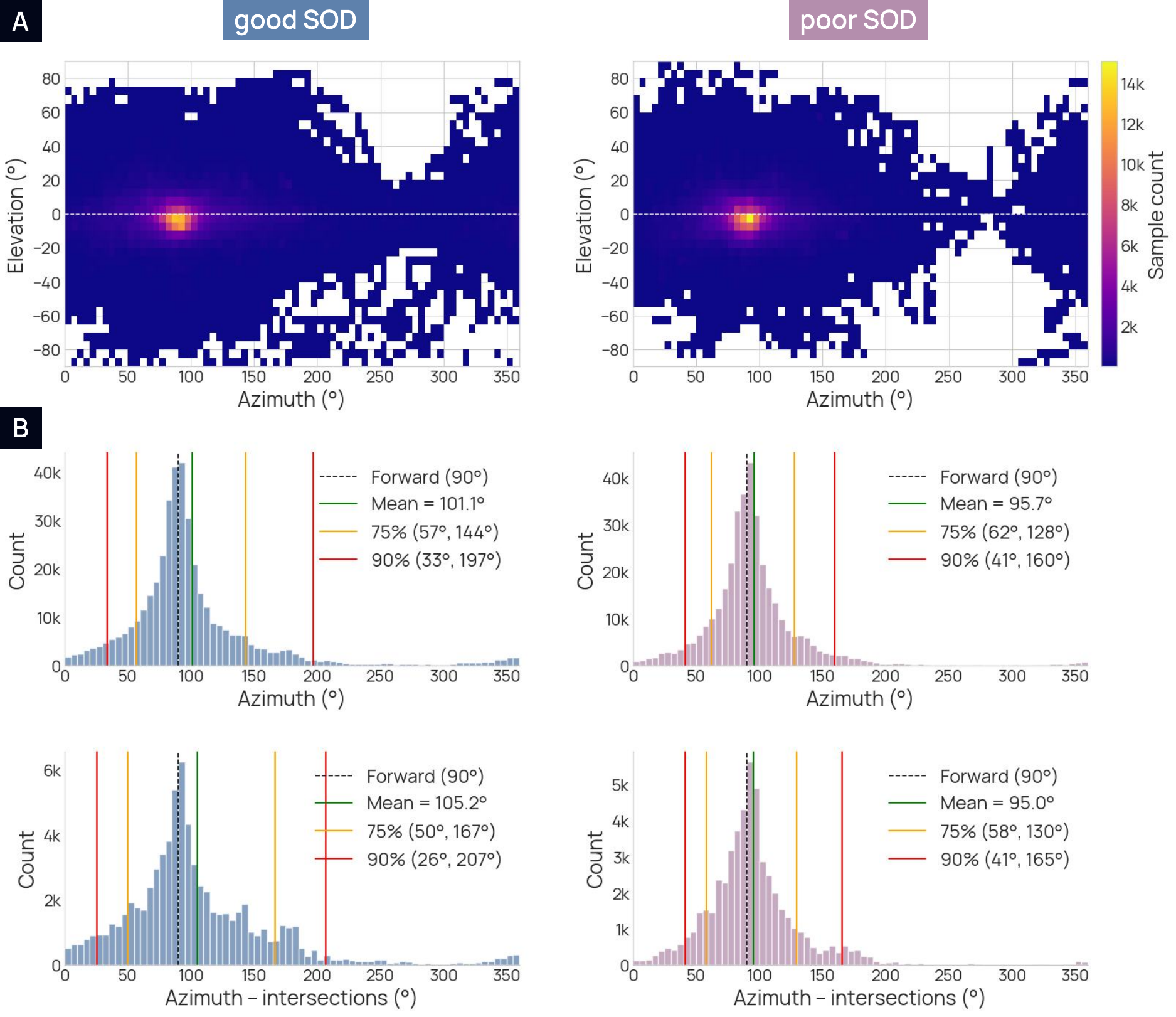}
    \caption{Gaze distribution profiles. (A): 2D heatmaps on equirectangular panoramas. (B): azimuth distributions with percentile boundaries used for VLM cropping extents.}
    \Description{Two panels. (A) shows pairs of 2D gaze heatmaps overlaid on equirectangular panoramas, comparing good and poor navigator groups, with warmer colors indicating higher gaze density. (B) shows azimuth distribution plots with shaded percentile boundaries marking the cropping extents used for the VLM pipeline.}
    \label{fig:gaze-distribution}
\end{figure}

\subsection{Results}
Our dataset contained approximately 1.65 million gaze samples across 19 participants, of which 804K were recorded on common routes.
The good SOD group contributed a mean of 76,626 samples per participant ($SD = 17,578$), and the poor SOD group contributed a mean of 75,169 ($SD = 15,099$). 
Mean total session duration was 75.4 minutes ($SD = 12.0$) for the good group and 71.5 minutes ($SD = 10.2$) for the poor group.

\subsubsection{Gaze Behavior}
On common routes, good navigators exhibited significantly greater total gaze
spread than poor navigators ($F(1,17) = 5.55$, $p < .05$, $\eta^2_p = .25$; $M_{\text{good}} = 0.13$, $SD = 0.03$; $M_{\text{poor}} = 0.11$, $SD = 0.03$).
This difference was driven by horizontal spread ($F(1,17) = 6.16$, $p < .05$,
$\eta^2_p = .27$), while vertical spread did not differ between groups
($F(1,17) = 0.87$, $p = .363$, $\eta^2_p = .05$).
In our analysis of fixation rate (\textit{i.e.,} how frequently participants locked onto a location) and mean fixation duration (\textit{i.e.,} how long each fixation lasted),
we found no significant group differences (rate: $\chi^2(1) = 0.08$,
$p = .772$; duration: $\chi^2(1) = 0.41$, $p = .520$). 
See~\autoref{fig:gaze-distribution}, ~\autoref{fig:gaze-patterns}.

\subsubsection{Head Rotation}
\begin{figure*}[t]
    \centering
    \includegraphics[width=\linewidth]{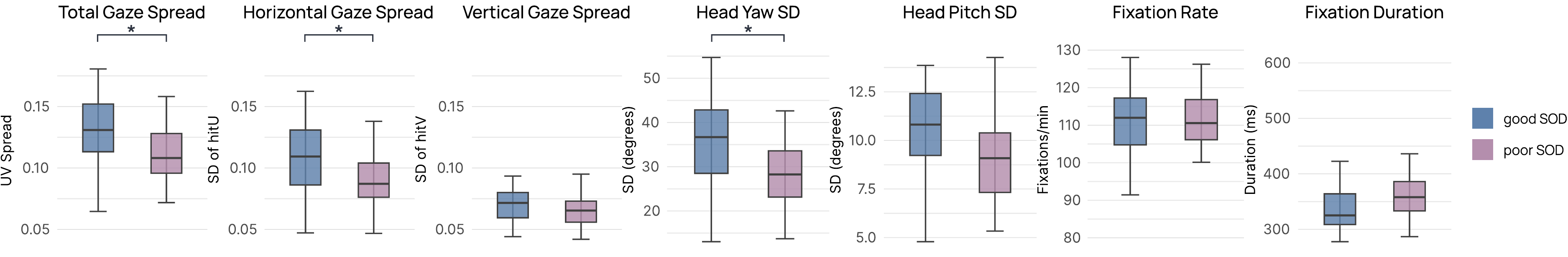}
    \caption{Gaze, fixation, and head movement by group. 
    Good navigators exhibited significantly greater total gaze spread, horizontal gaze spread and head yaw variability.
    Fixation patterns did not differ significantly between groups.
    }
    \Description{
    Seven side-by-side box plots comparing good SOD (blue) and poor SOD (pink) navigator groups. From left to right: Total Gaze Spread (UV Spread), Horizontal Gaze Spread (SD of hitU), and Vertical Gaze Spread (SD of hitV), each ranging from roughly 0.05 to 0.15; Head Yaw SD in degrees, ranging from about 20 to 50; Head Pitch SD in degrees, ranging from about 5 to 12.5; Fixation Rate in fixations per minute, ranging from about 80 to 130; and Fixation Duration in milliseconds, ranging from about 300 to 600. Asterisks mark significant differences for Total Gaze Spread, Horizontal Gaze Spread, and Head Yaw SD, where good navigators show higher values. Vertical Gaze Spread, Head Pitch SD, Fixation Rate, and Fixation Duration show overlapping distributions with no significant differences between groups.
    }
    \label{fig:gaze-patterns}
\end{figure*}

Good navigators turned their heads more than poor navigators. On common routes,
good navigators showed significantly greater head yaw variability
($F(1,17) = 6.59$, $p < .05$, $\eta^2_p = .28$; $M_{\text{good}} = 36.0^\circ$,
$SD = 9.7^\circ$; $M_{\text{poor}} = 28.2^\circ$, $SD = 6.8^\circ$) and head pitch
variability ($F(1,17) = 4.93$, $p < .05$, $\eta^2_p = .23$;
$M_{\text{good}} = 10.7^\circ$, $SD = 2.2^\circ$; $M_{\text{poor}} = 9.0^\circ$,
$SD = 2.1^\circ$).
See~\autoref{fig:gaze-patterns}.
Together, the gaze spread and head rotation findings suggest that good navigators survey a broader horizontal extent of the scene through both eye movements and physical head turns. 

\subsubsection{Route Retrace Performance}
In route retrace, good navigators chose the correct direction at intersections at a higher rate than poor navigators ($M_{\text{good}} = 75.5\%$, $SD = 29.3\%$; $M_{\text{poor}} = 57.5\%$, $SD = 27.8\%$). 
An analysis of variance based on mixed logistic regression indicated a statistically significant effect of group on intersection accuracy ($\chi^2(1, N=137) = 4.51$, $p < .05$). 
Retrace duration did not differ between groups
($M_{\text{good}} = 289.5\,\mathrm{s}$, $SD = 82.0$;
$M_{\text{poor}} = 312.0\,\mathrm{s}$, $SD = 84.9$; $F(1,9.2) = 1.01$, $p = .340$, $\eta^2_p = .10$). 
The retrace task served to maintain attentiveness throughout the session and to validate the SBSOD-based group assignment: despite each pair retracing a different route, the significant accuracy difference provides converging evidence that the groups captured meaningful differences in navigation ability.

\subsubsection{Verbal Data}
\label{sec:verbal-data-results}
We coded a total of 1,009 landmark mentions from good navigators and 1,170 from poor navigators across all transcripts (see Appendix \autoref{tab:qual-coding-results} for full results). 
Both groups predominantly referenced non-residential buildings, which accounted for 58.6\% and 52.0\% of mentions respectively. 
However, the groups diverged in what else they attended to. Good navigators disproportionately mentioned wayfinding signage (5.8\% vs. 1.8\%), while poor navigators more frequently referenced trivial details such as decorative elements (0.5\% vs. 3.5\%) and parts of buildings (0.6\% vs. 2.5\%).
The most striking differences emerged around transient and incidental features. 
Poor navigators referenced pedestrians (0.1\% vs. 1.9\%), parked bikes (0.3\% vs. 0.9\%), and generic signage that does not contain any location information (0.7\% vs. 2.6\%) far more frequently than good navigators. 
This is consistent with prior work~\cite{ishikawa_landmark_2012} that people with good SOD select more effective landmarks.

Inspecting frames with the most divergent gaze distributions between groups revealed consistent patterns. 
Good navigators attended to wayfinding signage (\autoref{fig:gaze-heatmaps-comparison}A) and permanent store signs such as names engraved in stone (\autoref{fig:gaze-heatmaps-comparison}B), while poor navigators fixated on menus and temporary store signage (\autoref{fig:gaze-heatmaps-comparison}C). 
Most notably, when walking perpendicular to the Tokyo Skytree (\autoref{fig:gaze-heatmaps-comparison}D), good navigators actively turned their heads to fixate on the tower as a global landmark, while poor navigators did not.

We also observed that the difference between groups was not only what caught their eye, but what they chose to articulate as useful for navigation.
When both groups gazed at the same campaign poster in Ginza, their verbal responses revealed different evaluative strategies. 
A01 (good SOD) dismissed it: \sayit{I don't need to mention it [campaign poster]. It will get removed later.} 
In contrast, A19 (poor SOD) said: \sayhi{Hopefully they don't change whatever that campaign is for next time I'm here.} 
This suggests that the gap lies not in perception but in an evaluative filter---good navigators actively suppress transient features from their mental representations, while poor navigators encode them despite awareness of their unreliability.

\section{\sysname}
\label{sec: system}

\begin{figure*}[b]
    \centering
    \includegraphics[width=\linewidth]{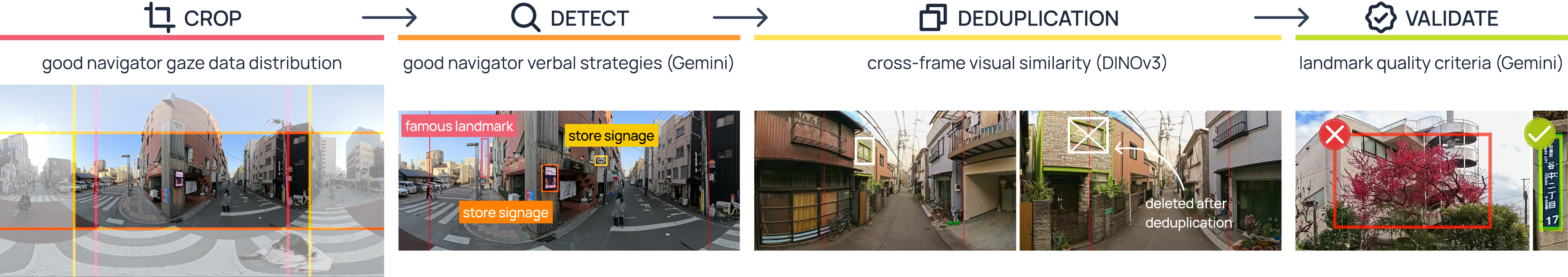}
    \caption{The landmark prediction pipeline: (1) gaze-informed cropping, (2) VLM-based detection using good navigators' verbal strategies, (3) embedding-based deduplication via DINOv3, and (4) second-pass validation.}
    \Description{A four-stage pipeline diagram flowing left to right. Stage 1 shows a panorama being cropped based on gaze distributions. Stage 2 shows the cropped image being processed by a VLM that outputs bounding boxes with landmark labels. Stage 3 shows embedding-based deduplication using DINOv3 to remove repeated detections across frames. Stage 4 shows a second-pass validation step that filters out false positives.}
    \label{fig:vlm-pipeline}
\end{figure*}

\sysname consists of two components: a landmark prediction pipeline and an MR navigation interface. 
Given a set of  $360^{\circ}$ images along a route, the prediction pipeline first crops each panorama to the forward-facing region where navigators concentrate their gaze, then passes the cropped images to a VLM that identifies navigation--relevant landmarks based on insights from our landmark attention study (\autoref{sec:landmark-attention-study}).
The navigation interface renders these landmarks as gaze-interactive overlays layered on top of conventional turn-by-turn directions, guiding users' attention towards visual features that skilled navigators select as effective clues.

\subsection{Landmark Prediction Pipeline}
\label{sec:pipeline}
 
Our landmark prediction pipeline translates findings from our landmark attention study into a scalable approach for identifying navigation-relevant landmarks in street-view imagery.
The pipeline operates in four stages: gaze-informed image cropping, VLM-based landmark detection, embedding-based deduplication, and second-pass validation (\autoref{fig:vlm-pipeline}).

\subsubsection{Gaze-Informed Image Cropping}
\label{sec:cropping}
Raw equirectangular panoramas span a full $360^{\circ} \times 180^{\circ}$ field of view, but navigators concentrate their gaze on a forward-facing subset. 
Passing full panoramas to a detection model dilutes predictions with irrelevant regions: sky, ground and areas behind the walker.
We therefore crop each panorama to where good navigators mostly look, using percentile boundaries from their gaze distributions (\autoref{fig:gaze-distribution}). 

Each panorama is first rotated to center the walking direction. 
At intersection frames, an additional half-turn rotation centers the crop on the bisector between the approaching and turning directions.
Based on the 90th-percentile azimuth interval, we crop to $180^{\circ}$ at
non-intersections (\autoref{fig:gaze-distribution}). 
At intersections we widen the crop to $270^{\circ}$ so that it covers both the approaching and the turning direction after the half-turn rotation, reflecting broader gaze spread at decision points.
Within the crop, a narrower guideline region taken from the 75th-percentile interval marks a preferred detection zone: $90^{\circ}$ at non-intersections and $120^{\circ}$ at intersections.
Vertical extent is fixed across both cases at the 95th-percentile elevation interval, which spans 0.5 in normalized $v$ coordinates. 

\subsubsection{VLM-Based Landmark Detection}
Given the cropped imagery, the pipeline identifies navigation-relevant landmarks using a VLM.
\newline
\textbf{Design Rationale.}
We initially explored fine-tuning a DeepGaze model~\cite{kummerer_modeling_2025} on good-navigator gaze heatmaps to predict visually salient regions.
However, the model achieved poor alignment with observed gaze (KL divergence of 1.62 on the Kinshicho common route); it often highlighted road surfaces and blank walls.
Moreover, our attention study showed that when the groups gazed at similar regions, the key distinction lies in which features they judged worth encoding for navigation (\autoref{sec:verbal-data-results})---a semantic distinction that a pixel-level saliency model cannot represent.
This motivated a VLM-based approach, where such distinctions can be expressed through natural language prompting.
\newline
\newline
\textbf{Prompt Design.}
The pipeline uses Google Gemini 3.1 Pro Preview~\cite{google_gemini_2026} for landmark detection.
The prompt instructs the model to identify landmarks drawn from priority-ranked categories from our qualitative codebook (\autoref{tab:qual-coding-results}), and to prioritize objects whose center falls within the guideline region while still permitting high-priority landmarks in the periphery.
Importantly, the prompt emphasizes \textit{relative distinctiveness} within each scene---a vending machine is a stronger anchor on a quiet residential street than on a busy commercial strip.
Non-intersection frames are processed in consecutive batches of five with a limit of two landmarks per image, while intersection frames are isolated into single-frame batches with a separate prompt allowing up to three landmarks. 
Bounding boxes tightly enclose the distinctive visual feature itself — \textit{e.g.,} a café's neon sign rather than the whole building.
Each detection is scored on color prominence, shape prominence, and semantic familiarity (\textit{i.e.,} how easily a person could describe the object), and returned as structured JSON via Gemini's \texttt{response\_schema} parameter, including a bounding box, description, category label, and navigational context.
See Appendix~\autoref{sec:detection-prompt} for details.

\subsubsection{Embedding-Based Deduplication}
The per-batch prompt deduplication reduces but does not eliminate duplicate detections across adjacent batches. 
For each detection, we crop the bounding box region from the rotated panorama and extract a feature embedding using the DINOv3 vision encoder~\cite{simeoni_dinov3_2025}. 
Within each route segment, we compare detections in nearby frames (±5 frame window) via cosine similarity of their L2-normalized CLS token embeddings. 
Because the pipeline processes each segment independently, we also compare detections across consecutive segment boundaries---matching the last frame of each segment against the first of the next. Detection pairs exceeding a similarity threshold of 0.70 are linked via union-find, and each group is pruned to retain only its earliest occurrence.

\subsubsection{Second-Pass Validation}
\label{sec:second-pass}
Finally, each remaining detection undergoes a second-pass validation. The detection's bounding box region is cropped from the source panorama (with a 1.5× buffer) and overlaid with a translucent rectangle marking the original detection area. 
This annotated crop is sent individually to Gemini along with the first-pass description and category label. 
A separate validation prompt (Appendix~\autoref{sec:validation-prompt}) instructs the model to verify that the described object is visible in the cropped region, that the category and bounding box refer to the same real-world object, that the object is permanent rather than transient, and that it is visually distinctive enough to serve as a landmark. 
The prompt includes an explicit exclusion list of generic municipal infrastructure---street lights, traffic signs, guardrails, utility poles---and temporary features such as banners and flags, reflecting the types of objects our qualitative findings showed poor navigators tend to over-encode. Invalid detections are excluded from the final landmark set. 

On average, the pipeline processes each panorama in eight seconds; for example, evaluation Route B (67 panoramas) took 9.2 minutes and cost USD \$1.77 in total.

\begin{figure*}[t]
    \centering
    \includegraphics[width=\linewidth]{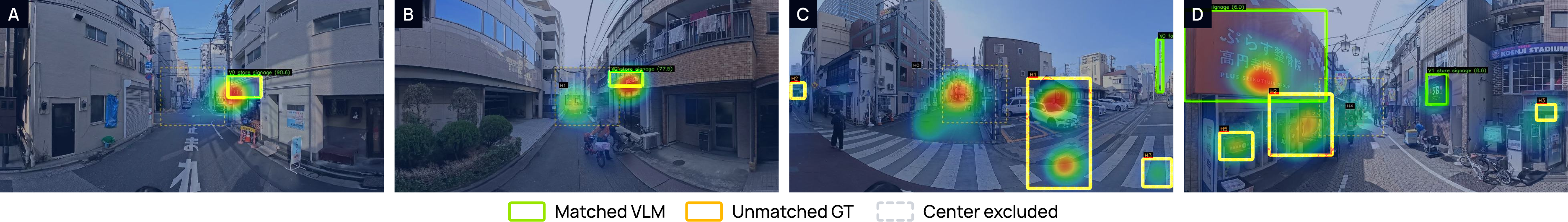}
    \caption{Examples of VLM detection results.
    In (A) and (B) VLM detections closely overlap with gaze hotspots. 
    In (C) and (D), unmatched gaze hotspots correspond to objects such as store interiors, parked cars, and sidewalks---features that attracted visual attention but never appeared in participants' verbal landmark descriptions.}
    \Description{Four street-view panoramas overlaid with gaze heatmaps and bounding boxes. Green boxes mark VLM detections that overlap with gaze hotspots; orange boxes mark gaze hotspots the VLM missed; dashed boxes mark the excluded center zone. In (A) and (B), VLM detections align closely with gaze concentrations on storefronts. In (C) and (D), unmatched orange boxes appear over store interiors, parked cars, and sidewalks where participants looked but never verbally referenced as landmarks.}
    \label{fig:vlm-results}
\end{figure*}

\begin{figure*}[b]
    \centering
    \includegraphics[width=\linewidth]{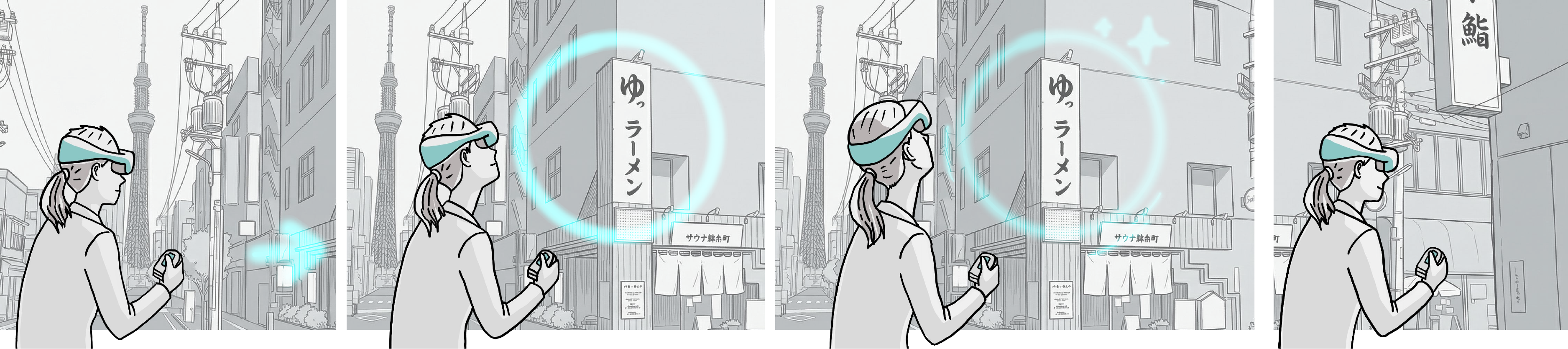}
    \caption{A walkthrough of LandmarkLens during navigation. As the user approaches an intersection, landmark highlights appear on the scene. A directional arrow prompts the user to look toward a highlight they haven't yet noticed. Once the user gazes at the landmark, the circle pulses and fades, and the user proceeds to the next panorama.}
    \Description{Four VR interface screenshots labeled A through D. (A) and (B) show the LandmarkLens condition with colored circle highlights on buildings and directional arrows pointing offscreen. (C) shows the scene-recognition task where a single street-view image is displayed with response options. (D) shows the route-choice task where a 360-degree intersection image is displayed with directional choice options (left, right, or straight).}
    \label{fig:landmarklens-interface-design}
\end{figure*}

\subsection{Evaluating the Landmark Guidance Pipeline}

\begin{table}[b]
    \centering
    \footnotesize
    \setlength{\tabcolsep}{3pt}
    
    \resizebox{\linewidth}{!}{%
    \begin{tabular}{lrrrrrrr}
    \toprule
    Area & Frames & \multicolumn{2}{c}{VLM} & \textbf{Precision} & \multicolumn{2}{c}{GT} & \textbf{Recall} \\
    \cmidrule(lr){3-4} \cmidrule(lr){6-7}
    & & Matched & Unmatched & & Matched & Unmatched & \\
    \midrule
    Ginza      & 105 & 123 & 19 & \textbf{86.6\%} & 75  & 64 & \textbf{54.0\%} \\
    Koenji     & 177 & 191 & 42 & \textbf{82.0\%} & 97 & 73 & \textbf{57.1\%} \\
    Nezu       & 149 & 85  & 30 & \textbf{73.9\%} & 32  & 55 & \textbf{36.8\%} \\
    Kinshicho  & 122 & 123 & 26 & \textbf{82.6\%} & 45  & 54 & \textbf{45.5\%} \\
    \bottomrule
    \end{tabular}%
    }
    
    \caption{VLM pipeline evaluation against ground-truth data.}
    \label{tab:pipeline-evaluation}
\end{table}

To assess whether the pipeline identifies objects that good navigators attend to, we compared VLM detections against ground-truth gaze heatmaps aggregated from the ten good-navigator participants across four common routes. 
We ran this evaluation on validated but non-deduplicated detections.
Since gaze data spans multiple consecutive frames for the same object, evaluating against deduplicated detections would artificially penalize recall.

For \textit{precision}---whether VLM detections fall where people actually looked---we compute each detection's density ratio: the share of total gaze captured by the bounding box divided by the box's area fraction, where values above 1.0 indicate the detection captures disproportionately more gaze than expected by chance. 
A detection is classified as matched if its density ratio $\geq 1.0$. For \textit{recall}---whether the VLM found what people looked at---we threshold the gaze heatmap at the 95th percentile of nonzero values and extract connected components as hotspot regions. 
Each hotspot is matched to VLM detections using IoU $\geq 0.1$ or IoMin $\geq 0.15$. 
A $\pm 1$ frame cross-matching pass accounts for the same landmark attracting gaze on adjacent frames. 
To avoid inflating both metrics with gaze attributable to forward path-monitoring rather than landmark recognition, we exclude detections and hotspots whose centroid falls within a center exclusion zone ($30^{\circ}$ wide × 15\% of panorama height) aligned with the travel direction.
\autoref{tab:pipeline-evaluation} reports results.
\autoref{fig:vlm-results} shows examples of the evaluation output. 
In cases of unmatched gaze hotspots, they correspond to objects such as store interiors, parked cars, and sidewalks—features that attracted visual attention but never appeared in verbal landmark descriptions.

\subsection{Guiding Landmark Attention in MR}
\textbf{Circle Highlights.}
As our goal is to direct gaze toward specific objects in the scene, we present landmarks as visual overlays rather than audio cues, an approach supported by prior findings that visual MR guidance reduces navigation errors and cognitive load compared to audio~\cite{zhao_effectiveness_2020}.
Each predicted landmark is presented as a high-saturation neon circle overlaid on the panorama at the predicted location (\autoref{fig:landmarklens-interface-design}). 
We chose circles over bounding boxes for two reasons. First, a circle's geometric simplicity stands out against the rectilinear urban environment. 
Second, unlike a tightly fitted bounding box that isolates an object, a circle encompasses both the landmark and its immediate surroundings, encouraging users to encode the object in the context of its environment. 
Circle color (magenta, orange, or yellow) is assigned based on the visual prominence scores from the VLM output, providing a subtle cue to the relative distinctiveness of each landmark. 
To balance engagement with distraction, each circle pulses when the user gazes at it and fades out after a two-second fixation. 
This gaze-to-dismiss interaction lightly gamifies the experience; users \say{collect} landmarks as they navigate, turning passive observation into an active visual scavenger hunt rewarding deliberate attention.

\noindent\textbf{Directional Arrows.} 
When a predicted landmark is outside the user's current field of view, a directional arrow appears at the edge of the display pointing toward it, guiding the user to rotate their head toward the landmark. 
Once the circle enters the field of view, the arrow parks at the display edge and remains visible until the circle is dismissed through the gaze interaction.

\noindent \textbf{Implementation.} 
We implemented \sysname as a VR navigation application in Unity, deployed on a Samsung Galaxy XR headset, extending the street-view navigation system from the landmark attention study with an additional landmark overlay layer.
\section{Preliminary User Evaluation}

\begin{figure*}[t]
    \centering
    \includegraphics[width=\linewidth]{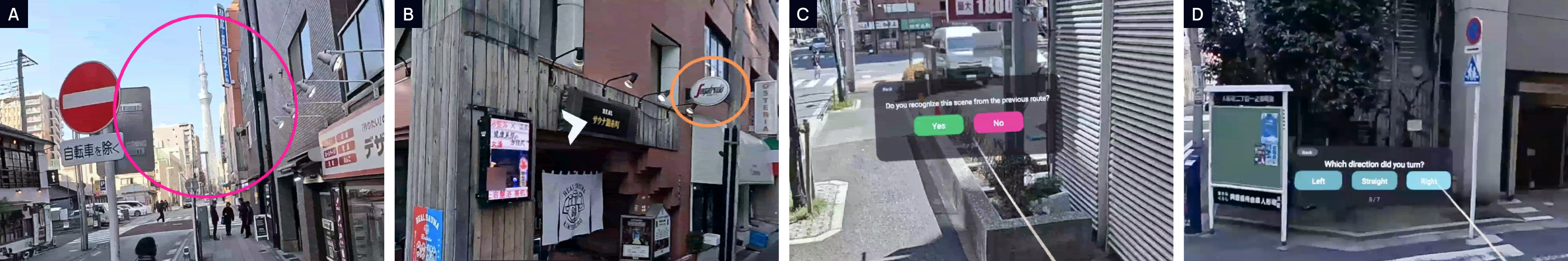}
    \caption{User Evaluation VR interfaces. (A–B) \sysname condition showing circle highlights and directional arrows. (C) Scene-recognition task. (D) Route-choice task.}
    \label{fig:evaluation-tasks}
    \Description{Four VR interface screenshots labeled A through D. (A) and (B) show the LandmarkLens condition with colored circle highlights on buildings and directional arrows pointing offscreen. (C) shows the scene-recognition task where a single street-view image is displayed with response options. (D) shows the route-choice task where a 360-degree intersection image is displayed with directional choice options (left, right, or straight).}
\end{figure*}

\begin{figure}[b]
    \centering
    \includegraphics[width=\linewidth]{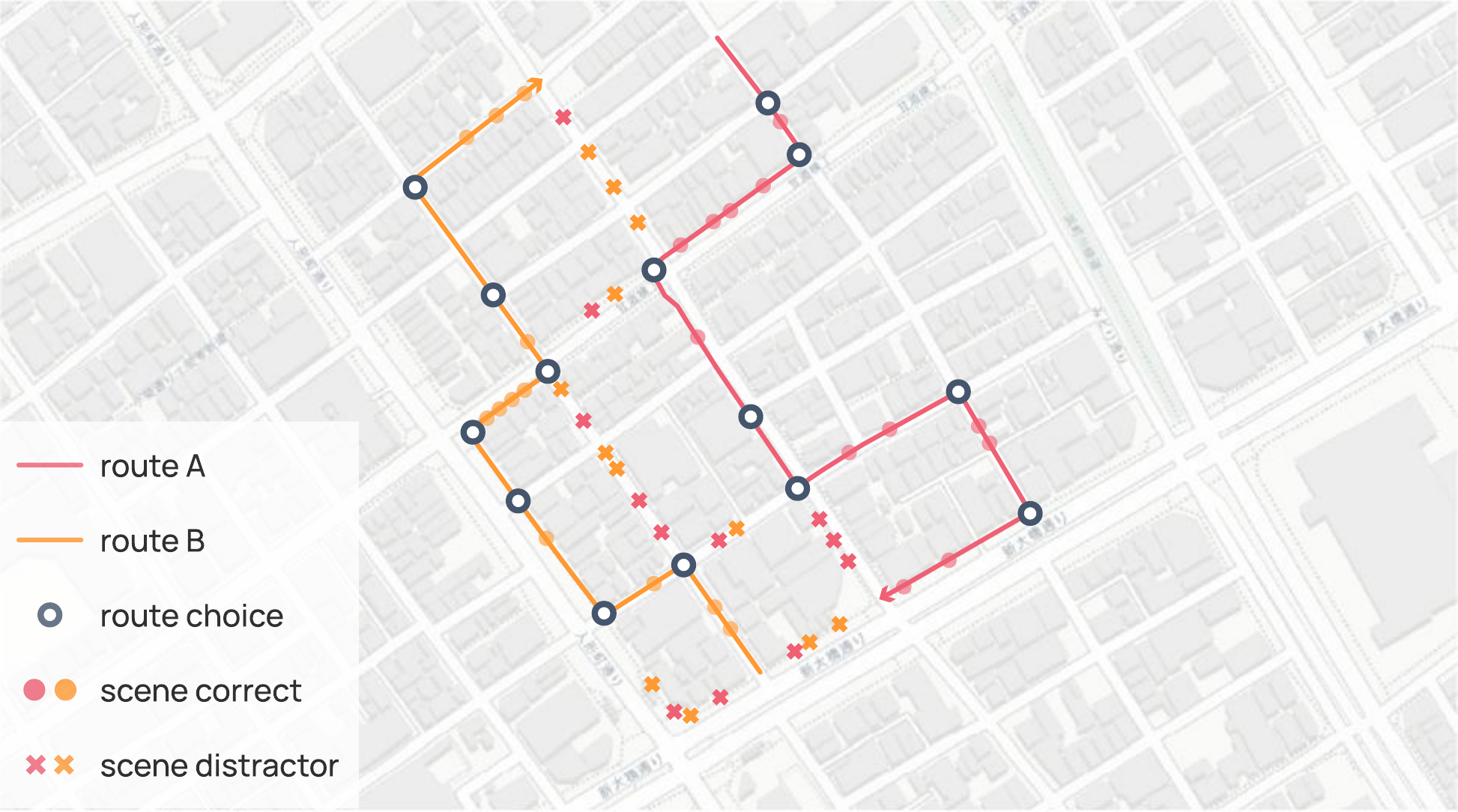}
    \caption{Evaluation routes and task image locations in Ningyocho. Filled circles: scene-recognition targets; crosses: distractors; open circles: route-choice intersections.}
    \Description{An overhead map of the Ningyocho neighborhood showing two evaluation routes drawn as paths through the street grid. Filled circles along the routes mark scene-recognition target locations, crosses mark distractor image locations on nearby streets, and open circles mark route-choice intersection locations.}
    \label{fig:evaluation-routes}
\end{figure}

To evaluate whether \sysname attention guidance supports spatial knowledge acquisition, we conducted a within-subjects study comparing \sysname against a turn-by-turn baseline.

\subsection{Participants}
We recruited eight participants (five women, three men; mean age = 27.4, $SD$ = 2.3), six from our previous landmark attention study and two through university mailing lists.
All participants scored $\leq 3.5$ on the SBSOD scale ($M = 2.8$, $SD = 0.4$).
Participants were compensated with a JPY \textyen5,000 gift card. 
See Appendix~\autoref{tab:evaluation-participants}.

\subsection{Study Design}
\textbf{Street-Level Imagery.}
To avoid familiarity effects for participants who had taken part in the landmark attention study, we collected new imagery from Ningyocho, a neighborhood that shares the same \textit{Mercantile Tokyo} archetype as Kinshicho.

\noindent\textbf{Route Selection.}
We selected two routes in Ningyocho (\autoref{fig:evaluation-routes}), balanced on the following criteria:
\begin{enumerate*}
    \item \textit{Length}: Route~A was 474.2\,m; Route~B was 410.9\,m.
    \item \textit{Structure}: Both routes contained six segments 
          and five turns, with a balanced number of left and right turns.
    \item \textit{Decision points}: Both included two additional straight-through intersections. 
    \item \textit{Visual anchors}: Both routes featured a mixture of commercial and residential buildings, with distinctive visual landmarks at turning intersections.
\end{enumerate*}
We applied the VLM landmark prediction pipeline (\autoref{sec: system}) to all frames.
The pipeline produced 49 highlights on Route~A (63 frames) and 58 on Route~B (67 frames), yielding a comparable density of ~0.8 highlights per frame.

\noindent \textbf{Task Design.}
Following~\cite{wen_instruction_2014}, we assessed participants' spatial knowledge using three tasks targeting landmark, route, and survey knowledge respectively. 
No time limit was imposed on any tasks.
In the \textit{scene-recognition task}, participants viewed 24 images in~VR, presented one at a time in random order, and indicated whether they recognized each scene from the exploration phase. 
Half of the images were taken from the explored route at non-intersection points; the other half were taken from nearby streets that participants had not explored (\autoref{fig:evaluation-routes}).
All route images were taken from frames where \sysname had placed at least one highlight.
In the \textit{route-choice task}, participants viewed seven $360^{\circ}$ images taken at intersections along the route and indicated which direction they had traveled during the exploration phase (left, right, or straight; \autoref{fig:evaluation-tasks}).
In the \textit{map-sketching task}, participants drew the explored route in as much detail as possible on a blank sheet of paper.

\subsection{Apparatus}
In both conditions, participants navigated the VR street-view system (\autoref{sec:landmark-attention-study}) by tilting the controller joystick, and text notifications at the bottom of the display indicated the required direction at each step (e.g., \textit{Go straight,} \textit{Turn right}).
In the \sysname condition, the system additionally displayed gaze-interactive landmark highlights generated by the VLM pipeline (\autoref{sec: system}).
Landmarks were presented as colored circles overlaid on the panorama at predicted locations, with an arrow indicator guiding participants toward circles outside their current field of view.
Participants dismissed each circle through a two-second fixation, after which it faded out.
Both conditions logged gaze and head-tracking data.
The scene-recognition and route-choice tasks were also administered in VR; the map-sketching task was completed on A4 paper.

\subsection{Procedure}
Each session lasted approximately 60 minutes across four phases: introduction, practice, route exploration with tasks, and post-study debrief.
We piloted with two participants to refine the procedure.

\noindent\textbf{Introduction.}
Each session began with an introduction to the study and informed consent. 
The moderator described the session structure and explained that after each route participants would be asked whether they recognized certain scenes, which direction they turned at intersections, and to sketch the route from memory.
 
\noindent \textbf{Practice.}
Participants completed two short practice routes to familiarize themselves with the system.
The first introduced basic navigation controls without landmark overlays; the second introduced the \sysname overlay. 
Participants were told that the highlighted circles indicated visual elements that might serve as helpful navigation cues, and were encouraged---but not required---to gaze at and dismiss each circle before advancing.
 
\noindent \textbf{Route Exploration and Tasks.}
Participants explored two routes, one with \sysname and one with the turn-by-turn baseline, in counterbalanced order. Route-to-condition assignment was also counterbalanced, yielding four counterbalance groups with two participants each.
After each route, participants completed the three spatial knowledge tasks in fixed order: map sketching, scene recognition, and route choice. 
This ordering follows~\cite{wen_instruction_2014} to minimize preceding tasks influencing subsequent ones.
Participants then rated seven Likert-scale items assessing confidence, spatial awareness, navigation independence, and perceived mental load.

\noindent\textbf{Post-Study.}
After completing both routes, participants answered Likert-scale questions on route memory, system usability, and perceived usefulness, followed by a short semi-structured interview.
 
\subsection{Analysis}
For task performance on scene recognition and route choice, we used \textit{mixed logistic regression} to model the proportion of correct responses for each task~\cite{hedeker_longitudinal_2006}.

For map sketching, we scored each sketch on route completeness
(segments drawn out of six), landmark count (number of landmarks freely recalled), and shape similarity. For shape similarity, we digitized each sketch as an SVG path, extracted vertex coordinates, and computed \textit{bidimensional regression}~\cite{tobler_bidimensional_1994}
between the sketch and the actual route geometry after optimal
translation, scaling, and rotation.  
Correlation values were transformed using Fisher's r-to-z transformation prior to statistical testing.
For ordinal Likert-scale responses (1--7), we used \textit{mixed ordinal logistic regression}~\cite{hedeker_random-effects_1994}.
For responses to open-ended questions, we focused on summarizing high-level themes.
One researcher developed themes through open coding~\cite{charmaz_constructing_2006} of interview transcripts and questionnaire responses.

\section{Findings}
On average, it took participants 6m 24s to finish the prototype condition ($SD$ = 2m 6s), and 3m 55s to finish the baseline condition ($SD$ = 2m 42s).
The longer completion time in the prototype condition reflects the gaze-to-dismiss interaction: although not required, most participants chose to dismiss all circles before advancing to the next panorama.
Below, we report quantitative results alongside participants' explanations.

\label{sec:findings}
\subsection{Task performance}

\begin{figure*}[t]
    \centering
    \includegraphics[width=\linewidth]{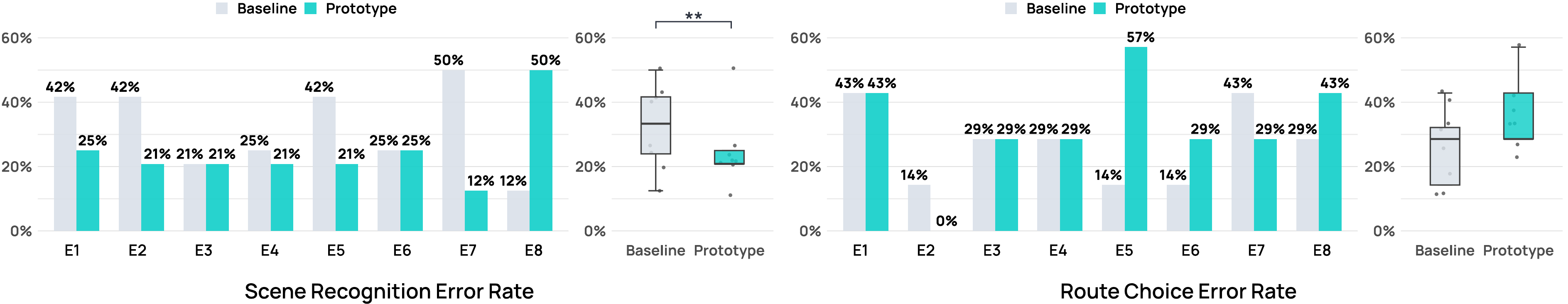}
    \caption{Left: Scene recognition error rate by participant and condition. LandmarkLens significantly improved accuracy over baseline (79.2\% vs. 64.9\%; p < .01).
    Right: Route-choice error rate per participant by condition. No significant difference (p = .535).}
    \Description{Two side-by-side charts. Left: a grouped bar chart or dot plot showing scene recognition accuracy by participant for the LandmarkLens and baseline conditions, with LandmarkLens averaging 79.2\% versus 64\% for baseline. Right: a chart showing route-choice errors per participant by condition, with no significant difference between conditions.}
    \label{fig:taskAB-results}
\end{figure*}

\begin{table}[b]
\centering
\footnotesize

\resizebox{\linewidth}{!}{%
\begin{tabular}{l cc cc cc}
\toprule
 & \multicolumn{2}{c}{\textbf{Route Completeness (\%)}} & \multicolumn{2}{c}{\textbf{Landmark Count}} & \multicolumn{2}{c}{\textbf{Shape Similarity ($r$)}} \\
\cmidrule(lr){2-3} \cmidrule(lr){4-5} \cmidrule(lr){6-7}
\textbf{PID} & Baseline & Prototype & Baseline & Prototype & Baseline & Prototype \\
\midrule
E1 & 50\%  & 33\%  & 6 & 6  & 0.96 & 0.94 \\
E2 & 100\% & 100\% & 0 & 1  & 0.96 & 0.98 \\
E3 & 100\% & 83\%  & 2 & 8  & 0.96 & 0.74 \\
E4 & 100\% & 100\% & 1 & 0  & 0.95 & 0.97 \\
E5 & 83\%  & 67\%  & 0 & 3  & 0.72 & 0.68 \\
E6 & 100\% & 83\%  & 4 & 7  & 0.73 & 0.92 \\
E7 & 100\% & 83\%  & 3 & 10 & 1.00 & 0.72 \\
E8 & 50\%  & 83\%  & 3 & 2  & 0.54 & 0.99 \\
\midrule
$Mdn$  & 100\% & 83\%  & 2.5 & 4.5 & 0.96 & 0.93 \\
$IQR$  & 50\%  & 17\%  & 3.0 & 5.5 & 0.23 & 0.24 \\
\bottomrule
\end{tabular}%
}

\caption{Map sketching results by participant and condition.}
\label{tab:map-sketching}
\end{table}

\subsubsection{Scene Recognition.}
Outlier analysis using z-score ($|z| > 2$) and IQR ($1.5 \times IQR$) methods identified one participant (E8) as an outlier on scene recognition; results are reported without this participant ($N = 7$).
Participants spent a similar amount of time on the scene recognition task across conditions, with a mean response time of 8.63 s ($SD = 3.94$) in the baseline condition and 9.47 s ($SD = 4.91$) in the prototype condition.

Participants recognized scenes more accurately with the prototype ($M=79.2\%, SD=4.2\%$) than with the baseline ($M=64.9\%, SD=11.1\%$).
An analysis of variance based on mixed logistic regression indicated a statistically significant effect of condition on scene recognition accuracy ($\chi^2(1,N=14)=8.36,p <.01$).
Participants attributed this improvement to the highlights drawing attention to features they would otherwise overlook. 
E1 said \sayit{the circles let me remember things I wouldn't normally notice,} 
and E6 remarked that \sayit{the tall buildings are things that I won't notice without the circles.}
E7 described the experience as \sayit{quite interesting\ldots it highlighted things that I would not usually pay attention to.}

The excluded participant, E8, offered insight into \textit{why} the highlights did not benefit her scene recognition. 
She reported that the circles shifted her attention away from her own encoding strategy: \sayit{I'm clearing the circles\ldots but I'm not really paying any attention to where I'm actually going or like where they are in relation to each other.} 
She explained that the highlights were most effective \sayit{when it aligned with the kind of stuff I noticed,} but because they were \sayit{not personalized,} they disrupted rather than supported her recall.

\begin{figure}[b]
    \centering
    \vspace{-1em}
    \includegraphics[width=\linewidth]{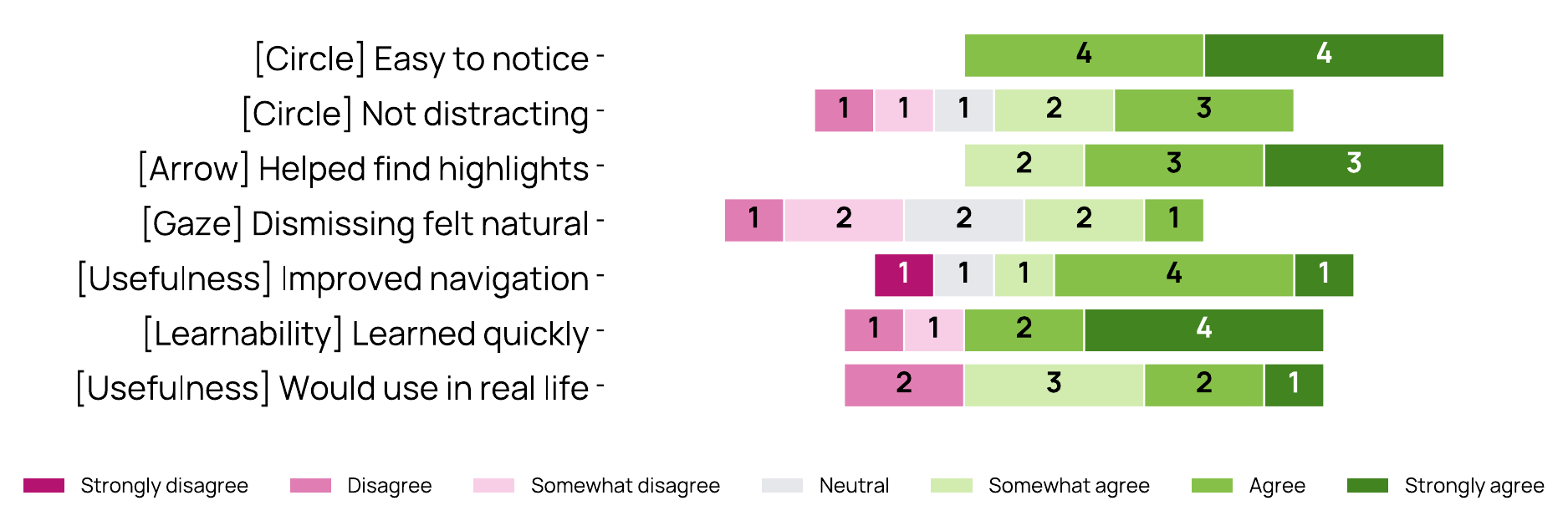}
    \caption{Prototype-specific Likert responses.}
    \Description{A diverging bar chart showing per-participant Likert-scale differences (prototype minus baseline) across seven post-route self-assessment items. Six of the seven items trend in favor of the prototype condition, with most differences falling on the positive side.}
    \label{fig:likert}
\end{figure}

\subsubsection{Route Choice.}
Participants' route choice accuracy was comparable across conditions 
($M_{\text{prototype}}$ = 67.9\%, $SD$ = 17.1\%; $M_{\text{baseline}}$ = 73.2\%, $SD$ = 11.4\%). 
An analysis of variance based on mixed logistic regression indicated no significant effect of condition on route choice accuracy ($\chi^2(1, N=16) = 0.39$, $p = .535$).
Response times for the route choice task were comparable between conditions, with a mean of 9.49 s ($SD = 5.26$) in the baseline condition and 9.79 s ($SD = 5.97$) in the prototype condition.

A recurring theme in the interviews was that the volume of highlights could interfere with route encoding. 
E5 mentioned that \sayit{too many highlights can make me [\ldots] only remember those surroundings, but not the route,} and E8 similarly noted that \sayit{because the circles were constant, it was harder for me to remember what landmarks were associated with some kind of directional change.} 

\subsubsection{Map Sketching.}
Map sketching performance was assessed along three dimensions:
route completeness, landmark count, and shape similarity.
\autoref{tab:map-sketching} reports per-participant results.

Participants recalled nearly twice as many landmarks with the
prototype ($Mdn = 4.5$, $IQR = 5.5$) as with
the baseline ($Mdn = 2.5$, $IQR = 3.0$). 
An analysis of variance based on mixed Poisson regression indicated
a statistically significant effect of condition on landmark count
($\chi^2(1, N=16) = 5.80$, $p = .016$).
Route completeness was comparable across conditions ($Mdn_{\text{baseline}} = 100\%$, $IQR = 50\%$;
$Mdn_{\text{prototype}} = 83\%$, $IQR = 17\%$); 
an analysis of variance based on mixed logistic regression indicated no significant difference ($\chi^2(1, N=16) = 0.77$, $p = .381$).
Structural accuracy did not differ significantly between
conditions. 
The median bidimensional correlation was comparable
across baseline ($Mdn_{\text{baseline}} = .96$) and prototype
($Mdn_{\text{prototype}} = .93$). 
A paired-samples t-test on
Fisher z-transformed correlations indicated no significant
difference ($t(7) = -0.12$, $p = .911$, $d = -0.04$).
Taken together, the map sketching results suggest that the
prototype primarily enhanced landmark encoding---participants
recalled nearly twice as many landmarks---without compromising
the overall structural accuracy of their mental representations.
See Appendix~\autoref{fig:evaluation-sketch} for full participant sketches and analysis.

\begin{figure}[b]
    \centering
    \vspace{-1em}
    \includegraphics[width=\linewidth]{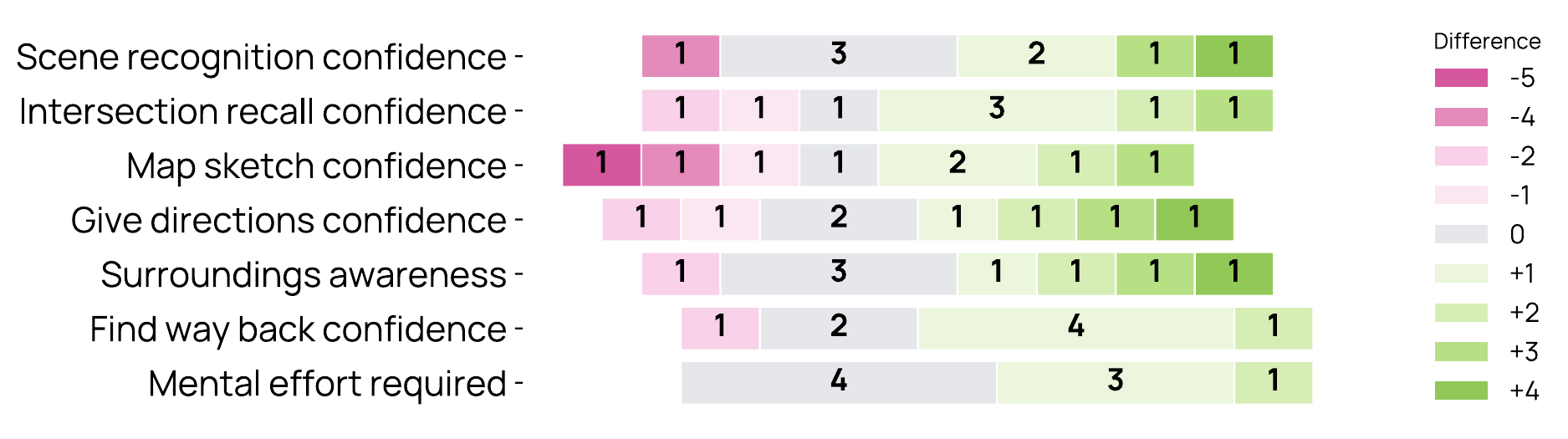}
    \caption{Per-participant Likert-scale differences (prototype-baseline) across post-route self-assessment items. 
    Six of seven items trended in favor of prototype.}
    \Description{A horizontal bar chart or Likert response distribution showing participants' ratings on prototype-specific questions, including ease of noticing circles, helpfulness of directional arrows, ease of learning, navigation improvement, naturalness of gaze-based dismissal, distraction level, and willingness to use the system with AR glasses in real life.}
    \label{fig:likert-diff}
\end{figure}

\subsection{Subjective Experience}

\subsubsection{Post-Route Self-Assessment.}
Figure~\ref{fig:likert-diff} shows the distribution of per-participant differences (prototype $-$ baseline) across seven post-route
Likert-scale items. 
Analyses of variance based on mixed ordinal
logistic regression~\cite{hedeker_random-effects_1994} indicated
no statistically significant effects of condition on any
individual item. 
Descriptively, however, six of seven items trended
in favor of the prototype, with participants reporting higher
confidence in scene recognition, intersection recall,
surroundings awareness, and finding their way back, as well as
lower mental effort. Map sketch confidence was the only item
without a clear directional pattern.

\subsubsection{Prototype Feedback.}
When asked which route they would remember better if they navigated it again the next day, six of eight participants chose the prototype route.
\autoref{fig:likert} summarizes participants'
responses to prototype-specific questions. The landmark highlight
circles were easy to notice ($Mdn = 6.5$), and the
directional arrows were well-received, with all participants
rating them positively for helping locate highlighted features
($Mdn = 6.0$). 
Participants generally found the system easy to learn ($Mdn = 6.5$) and most agreed it improved their navigation experience ($Mdn = 6.0$).
Opinions were more mixed on whether gaze-based dismissal felt natural ($Mdn = 4.0$) and whether the circles were distracting ($Mdn = 3.0$).
Willingness to use such a system with AR glasses in real life was generally positive ($Mdn = 5.0$).

Quantitative ratings were largely positive, and interviews revealed enthusiasm for pushing the system toward more \sayit{semantically meaningful and personalized} (E8) highlights. 
E3, who has a background in architecture, was engaged by \textit{buildings with special façades}, illustrating how personal experience shapes what constitutes a useful landmark. 
E7 suggested adding contextual information such as a building's history, and E1 proposed color-coding circles by building use. 
For participants with limited Japanese reading proficiency (E4 and E8), highlights on Japanese signage were difficult to retain---\sayit {I can't remember the signs I can't read} (E4).
\section{Discussion}
\textbf{Guided Landmark Attention Partially Supports Spatial Knowledge Acquisition.}
\sysname improved scene recognition but had more modest effects on route choice and map sketching. Following Sorrows and Hirtle's taxonomy~\cite{sorrows_nature_1999}, our VLM pipeline primarily detects visual landmarks, which directly support scene-level recognition. 
However, route and survey knowledge depend more on structural and semantic landmarks, which our pipeline does not yet target. 
Participant feedback corroborated this: highlights were most useful at turns but overwhelming along straight segments (E3, E5, E8), and landmarks with unreadable text were impossible to remember (E4). 
Still, because landmark knowledge is the foundation for route and survey knowledge~\cite{siegel_development_1975}, 
an open question is whether repeated landmark-level gains could, over time, scaffold the acquisition of route and survey knowledge~\cite{siegel_development_1975}.
Investigating this longitudinal effect is a promising future direction.

\noindent\textbf{VLMs as A Viable Approach for Landmark Prediction.}
VLMs are well-suited because the gap between good and poor navigators is an evaluative filter, not a perceptual one—good navigators see the same features but withhold irrelevant ones from their reports.
This filter maps naturally onto VLM instruction-following: we encoded good navigators' qualitative patterns directly into prompts, something purely vision-based saliency models cannot do. 
The approach also extends naturally: route structure (GPS, intersection topology) could enable context-sensitive highlighting at decision points, and because selection is prompt-mediated, it could personalize to language proficiency or area familiarity without retraining.

\noindent\textbf{Individual Variability and Personalization.} Our pipeline presents the same landmarks to all users, but we did observe within-group differences, where individuals within the good SOD group attended to different landmarks in the same environment (Appendix~\autoref{fig:individual-gaze}). 
This aligns with the concept of cognitive landmarks, which hold personal meaning or relevance to an individual~\cite{sorrows_nature_1999}.
\sysname does not yet account for these individual differences, and personalization based on user preference, language proficiency, and area familiarity is a natural next step.

\noindent\textbf{Ecological Validity of VR-Based Studies.}
Both our landmark attention study and the user evaluation were conducted in VR using static $360^\circ$ street-view imagery rather than real-world navigation.
Prior work has shown that high-fidelity VR combined with naturalistic movement control can narrow the gap between VR and real-world gaze behavior~\cite{drewes_gaze_2021}.
However, we acknowledge that our environment lacked dynamic scene elements such as moving vehicles and pedestrians, vestibular cues from physical locomotion, and peripheral visual information beyond the headset's field of view, all of which may influence gaze behavior and spatial learning during real-world navigation.
While many systems intended for AR have been designed and evaluated in VR as a first step (\textit{e.g.,} ~\cite{cheng_towards_2022}), future work should evaluate \sysname with real-world navigation.

\noindent\textbf{Future Work.}
\sysname was evaluated in VR but designed for AR smart glasses—the gaze-interactive overlays (dwell-based dismissal, directional arrows) transfer directly to head-worn displays. 
Real-world deployment introduces challenges (lighting variation, dynamic occlusion, inference latency) but also richer spatial encoding through actual locomotion.
Our evaluation was also limited in scope: the final study involved eight participants, and all routes were drawn within Tokyo. 
Evaluation with larger samples and in other regions with different landmark densities or urban layouts would help establish the generalizability of both our pipeline and the attention guidance approach.
More broadly, this work points toward a design paradigm for next-generation AR navigation: tools that guide users to their destination may support users in building their capacity to navigate independently, helping break the cycle in which GPS dependence erodes spatial ability~\cite{ishikawa_satellite_2019, ruginski_gps_2019}.
\section{Conclusion}
This paper presented \sysname, an MR navigation system that guides people with poor sense of direction toward landmarks that skilled navigators rely on. 
A landmark attention study with 20 participants revealed that good navigators scan a broader horizontal extent of the scene and selectively articulate durable wayfinding cues, while poor navigators disproportionately reference transient features---a gap rooted in evaluative filtering rather than perception.
We translated these findings into a VLM-based landmark prediction pipeline and gaze-interactive overlay system. 
A preliminary evaluation with eight poor-SOD participants showed that \sysname significantly improved scene recognition accuracy and nearly doubled landmark recall in map sketches compared to a turn-by-turn baseline, while route and survey knowledge remained comparable. 
Participant feedback highlighted that landmarks were most valued at decision points and that highlights should be semantically meaningful and personalized, suggesting that future iterations should adopt context-sensitive highlighting that concentrates cues near intersections and adapts to individual familiarity and preferences.
\begin{acks}
We thank the reviewers for their insightful comments and our study participants for their time. We also thank Maurício Sousa,  Xia Ding, I-Chao Shen, Hongbo Zhao, Zhongyi Zhou, Yuan Li, Jaewook Lee, Quentin Becker, Maria Larsson, and Xinyue Gui for their feedback and support throughout this project. This work was supported by the Institute for AI and Beyond of the University of Tokyo, by the Japan Science and Technology Agency (JST) under the Adopting Sustainable Partnerships for Innovative Research Ecosystem (ASPIRE) program, Grant Number JPMJAP2401, and by NSF Grant \#2411222.
\end{acks}

\newpage
\bibliographystyle{ACM-Reference-Format}
\bibliography{references}

\appendix
\onecolumn
\label{sec:appendix}
\section{Landmark Attention Study Materials \& Additional Analysis}
\begin{figure*}[ht]
    \centering
    \includegraphics[width=\linewidth]{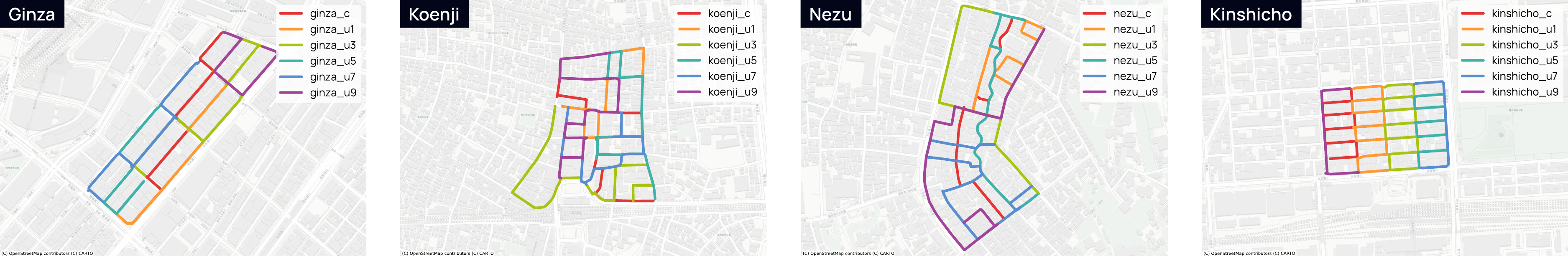}
    \caption{Route selections for landmark attention study. Each neighborhood included one common route (all 20 participants) and ten unique routes (one per participant).}
    \label{fig:attention-study-routes}
    \Description{An overhead map showing route layouts across the four Tokyo neighborhoods (Ginza, Koenji, Nezu, and Kinshicho). Each neighborhood panel shows one common route highlighted distinctly and ten unique routes, illustrating the spatial distribution and coverage of walking paths used in the landmark attention study.}
\end{figure*}

\begin{table*}[ht]
  \centering
  \small
  \setlength{\tabcolsep}{3pt}
  \setlength{\aboverulesep}{0pt}
  \setlength{\belowrulesep}{0pt}
  \setlength{\extrarowheight}{2.5pt}
  \renewcommand{\arraystretch}{1.15}
  \rowcolors{2}{white}{rowgray}
    \begin{tabular}{@{}M{0.045\linewidth} M{0.03\linewidth} M{0.06\linewidth} M{0.08\linewidth} M{0.062\linewidth} M{0.235\linewidth} R{0.06\linewidth} R{0.035\linewidth} R{0.035\linewidth} R{0.035\linewidth} R{0.035\linewidth} R{0.035\linewidth} R{0.035\linewidth} R{0.035\linewidth} R{0.035\linewidth}@{}}
    \textbf{PID} & \textbf{Age} & \textbf{Gender} & \textbf{SBSOD} & \textbf{Group} & \textbf{Languages} & \textbf{JP Read} & \textbf{R1} & \textbf{R2} & \textbf{R3} & \textbf{R4} & \textbf{R5} & \textbf{R6} & \textbf{R7} & \textbf{R8} \\
    \toprule
    A01 & 29 & F & 5.2 & Good & Chinese, English & 4 & 10 & 6 & 1 & 1 & 1 & 1 & 1 & 1 \\
    A02 & 26 & M & 5.3 & Good & Chinese, English, Japanese & 4 & 1 & 1 & 1 & 1 & 6 & 4 & 1 & 1 \\
    A03 & 28 & M & 5.2 & Good & Chinese, Japanese, English, German & 4 & 2 & 6 & 3 & 3 & 1 & 2 & 1 & 1 \\
    A04 & 23 & M & 6.9 & Good & Chinese, English & 5 & 1 & 1 & 1 & 1 & 1 & 1 & 5 & 5 \\
    A05 & 27 & M & 6.8 & Good & Chinese, Japanese, English & 5 & 2 & 2 & 1 & 1 & 1 & 1 & 1 & 1 \\
    A06 & 29 & M & 5.5 & Good & Chinese, English & 3 & 2 & 3 & 3 & 3 & 5 & 5 & 2 & 1 \\
    A07 & 25 & M & 5.7 & Good & Chinese, English, Japanese & 5 & 2 & 2 & 2 & 2 & 1 & 1 & 1 & 1 \\
    A08 & 30 & M & 5.7 & Good & French, English & 2 & 1 & 2 & 1 & 1 & 1 & 1 & 6 & 6 \\
    A09 & 29 & F & 6.6 & Good & Chinese, English, Japanese & 5 & 4 & 5 & 1 & 1 & 1 & 1 & 1 & 1 \\
    A10 & 23 & M & 6.7 & Good & English, Hindi & 3 & 4 & 1 & 1 & 1 & 5 & 1 & 1 & 1 \\
    \midrule
    A11 & 28 & F & 3.4 & Poor & Chinese, English & 3 & 6 & 7 & 3 & 2 & 2 & 2 & 2 & 2 \\
    A12 & 27 & M & 2.9 & Poor & Japanese, English & 5 & 1 & 1 & 1 & 1 & 8 & 8 & 1 & 1 \\
    A13 & 25 & F & 2.1 & Poor & Chinese, Japanese, English & 4 & 5 & 3 & 1 & 1 & 1 & 1 & 1 & 1 \\
    A14 & 39 & F & 3.4 & Poor & Swedish, English, Japanese, German, French & 3 & 3 & 2 & 10 & 10 & 3 & 3 & 8 & 8 \\
    A15 & 27 & M & 3.3 & Poor & English, Hindi & 5 & 4 & 3 & 2 & 2 & 8 & 7 & 1 & 1 \\
    A16 & 30 & F & 2.8 & Poor & Chinese, English, Japanese & 4 & 1 & 1 & 1 & 1 & 2 & 1 & 1 & 1 \\
    A17 & 28 & M & 3.4 & Poor & English & 2 & 1 & 1 & 3 & 4 & 1 & 1 & 1 & 1 \\
    A18 & 28 & M & 2.3 & Poor & Chinese, English & 2 & 3 & 1 & 1 & 1 & 1 & 1 & 3 & 2 \\
    A19 & 25 & F & 2.3 & Poor & English & 2 & 3 & 2 & 1 & 1 & 1 & 1 & 2 & 1 \\
    A20 & 25 & F & 2.9 & Poor & Chinese, English & 4 & 1 & 2 & 1 & 1 & 5 & 4 & 6 & 1 \\
  \end{tabular}
  \vspace{1em}
  \caption{Participant demographics, spatial ability (SBSOD), and self-reported
    familiarity with the route walked in each position R1--R8, separated into good
    and poor navigator groups. Route positions map to specific routes in~\autoref{tab:route-order};
    familiarity was rated from 1 (never been here before) to 10 (walk here regularly).
    Japanese reading levels range from 1 (not at all) to 5 (very well).}
  \Description{Table of twenty study participants split into a good navigator group and a poor navigator group, listing age, gender, SBSOD score, languages spoken fluently, self-rated Japanese reading level, and familiarity ratings for each of the eight study routes.}
  \label{tab:attention-study-participants}
\end{table*}

\begin{table*}[t]
  \centering
  \small
  \setlength{\tabcolsep}{4pt}
  \setlength{\aboverulesep}{0pt}
  \setlength{\belowrulesep}{0pt}
  \setlength{\extrarowheight}{2.5pt}
  \renewcommand{\arraystretch}{1.15}
  \rowcolors{2}{white}{rowgray}
  \begin{tabular}{@{}M{0.070\linewidth} M{0.0995\linewidth} M{0.0995\linewidth} M{0.105\linewidth} M{0.0995\linewidth} M{0.0995\linewidth} M{0.0995\linewidth} M{0.0995\linewidth} M{0.0995\linewidth}@{}}
    \textbf{PIDs} & \textbf{R1} & \textbf{R2} & \textbf{R3} & \textbf{R4} & \textbf{R5} & \textbf{R6} & \textbf{R7} & \textbf{R8} \\
    \toprule
    A01/A11 & ginza\_c      & ginza\_u1     & koenji\_u1     & koenji\_c      & nezu\_c        & nezu\_u1       & kinshicho\_u1 & kinshicho\_c  \\
    A02/A12 & koenji\_u2    & koenji\_c     & kinshicho\_c   & kinshicho\_u2  & ginza\_u2      & ginza\_c       & nezu\_c       & nezu\_u2      \\
    A03/A13 & nezu\_c       & nezu\_u3      & ginza\_u3      & ginza\_c       & kinshicho\_c   & kinshicho\_u3  & koenji\_u3    & koenji\_c     \\
    A04/A14 & kinshicho\_u4 & kinshicho\_c  & nezu\_c        & nezu\_u4       & koenji\_u4     & koenji\_c      & ginza\_c      & ginza\_u4     \\
    A05/A15 & ginza\_u5     & ginza\_c      & koenji\_c      & koenji\_u5     & nezu\_u5       & nezu\_c        & kinshicho\_c  & kinshicho\_u5 \\
    A06/A16 & koenji\_c     & koenji\_u6    & kinshicho\_u6  & kinshicho\_c   & ginza\_c       & ginza\_u6      & nezu\_u6      & nezu\_c       \\
    A07/A17 & nezu\_u7      & nezu\_c       & ginza\_c       & ginza\_u7      & kinshicho\_u7  & kinshicho\_c   & koenji\_c     & koenji\_u7    \\
    A08/A18 & kinshicho\_c  & kinshicho\_u8 & nezu\_u8       & nezu\_c        & koenji\_c      & koenji\_u8     & ginza\_u8     & ginza\_c      \\
    A09/A19 & ginza\_u9     & ginza\_c      & koenji\_c      & koenji\_u9     & nezu\_u9       & nezu\_c        & kinshicho\_c  & kinshicho\_u9 \\
    A10/A20 & koenji\_c     & koenji\_u10   & kinshicho\_u10 & kinshicho\_c   & ginza\_c       & ginza\_u10     & nezu\_u10     & nezu\_c       \\
  \end{tabular}
  \caption{Route presentation order (R1--R8) for each of the ten counterbalanced orders. Route labels match the legends in~\autoref{fig:attention-study-routes}.
  Each order was walked by one good navigator (A01--A10) and one poor navigator (A11--A20). Neighborhoods were presented in blocks of two routes, alternating which of the pair was the common route.}
  \Description{Table of ten counterbalanced route orders. Each row lists a pair of participant IDs, one good navigator and one poor navigator, followed by the eight routes they walked in order. Routes are named by neighborhood (Ginza, Koenji, Nezu, Kinshicho) and by whether they were the common route or a unique route.}
  \label{tab:route-order}
\end{table*}

\begin{figure*}[h]
    \centering
    \includegraphics[width=\linewidth]{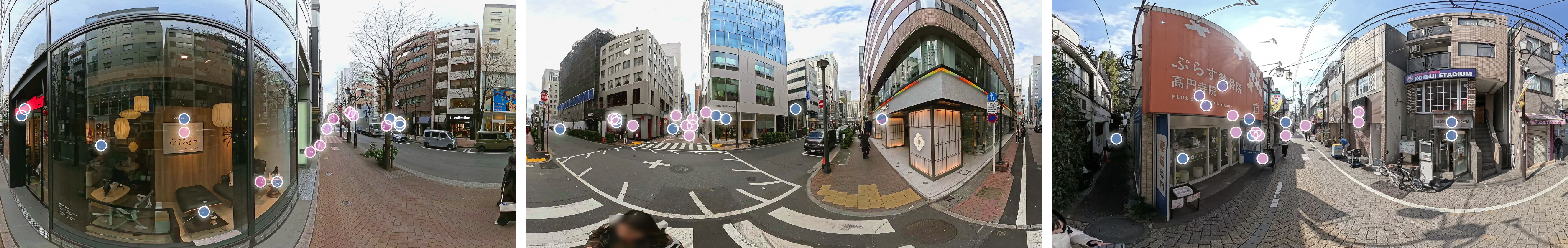}
    \caption{Gaze fixation centers for three frames with the highest within-group dispersion among good navigators. 
    Each dot represents one participant's primary fixation location.
    Teal dots represent good SOD participants; pink dots represent poor SOD participants.}
    \label{fig:individual-gaze}
    \Description{Three street-view panoramas of Tokyo scenes, each with teal and pink dots scattered across the image. Each dot marks one participant's primary fixation location. Blue dots (good SOD) and pink dots (poor SOD) are spread across different features in each frame, showing high within-group dispersion among good navigators across storefronts, signage, and street elements.}
\end{figure*}

\begin{figure*}[h]
    \centering
    \includegraphics[width=\linewidth]{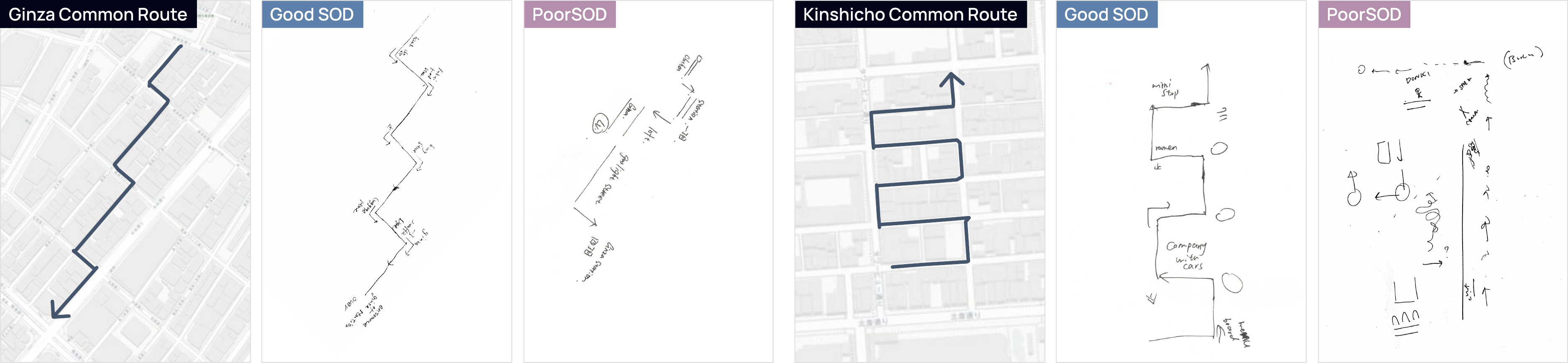}
    \caption{
    Example sketch maps from landmark attention study for two common routes.
    Each triplet shows the actual route geometry, a sketch from a good SOD participant, and a sketch from a poor SOD participant.
    Good SOD participants' sketches tend to preserve route geometry and relative spatial relationships.}
    \Description{Six panels arranged as two triplets. Each triplet shows a Ginza common route: the actual route geometry on a satellite map (left), a hand-drawn sketch from a good SOD participant (center), and a sketch from a poor SOD participant (right). The good SOD sketches closely preserve the route's turn angles and segment proportions with labeled landmarks, while the poor SOD sketches show less accurate spatial relationships and fewer or less organized landmark annotations.}
    \label{fig:attention-study-map-sketching}
    \Description{Six panels arranged as two triplets. Each triplet shows a Ginza common route: the actual route geometry on a satellite map (left), a hand-drawn sketch from a good SOD participant (center), and a sketch from a poor SOD participant (right). The good SOD sketches closely preserve the route's turn angles and segment proportions with labeled landmarks, while the poor SOD sketches show less accurate spatial relationships and fewer or less organized landmark annotations.}
\end{figure*}

\begin{table*}[h]
  \centering
  \small
  \setlength{\tabcolsep}{4pt}
  \setlength{\aboverulesep}{0pt}
  \setlength{\belowrulesep}{0pt}
  \setlength{\extrarowheight}{2.5pt}
  \renewcommand{\arraystretch}{1.15}
  \begin{tabular}{@{}L{0.135\linewidth} M{0.16\linewidth} M{0.40\linewidth} R{0.055\linewidth} R{0.045\linewidth} R{0.055\linewidth} R{0.045\linewidth}@{}}
    \textbf{Category} & \textbf{Tag} & \textbf{Example} & \textbf{Good} & \textbf{\%} & \textbf{Poor} & \textbf{\%} \\
    \toprule
    \multirow{10}{=}{Buildings \& Structures}
      & \sh non-residential buildings & \sh restaurants, shops, bars, salon, school, clinic, bank, senior home, community center, gallery, post office & \sh 591 & \sh 58.6 & \sh 608 & \sh 52.0 \\
      & residential buildings & apartments, single family house & 46 & 4.6 & 63 & 5.4 \\
      & \sh convenience store & \sh 7-11, FamilyMart, Lawson, Mini Stop & \sh 17 & \sh 1.7 & \sh 29 & \sh 2.5 \\
      & parts of a building & roof, stairs, wall, structurally necessary elements & 6 & 0.6 & 29 & 2.5 \\
      & \sh structures & \sh Tokyo Tower, Tokyo Skytree & \sh 6 & \sh 0.6 & \sh 8 & \sh 0.7 \\
      & items inside of a store/house & --- & 6 & 0.6 & 3 & 0.3 \\
      & \sh decorative details & \sh doorbell, potted plants, façade decoration, tiles, structurally unnecessary elements & \sh 5 & \sh 0.5 & \sh 41 & \sh 3.5 \\
    \midrule
    \multirow{9}{=}{Urban Infrastructure}
      & \sh road features & \sh crosswalks, T-intersection, four-way intersection, Y split & \sh 47 & \sh 4.7 & \sh 25 & \sh 2.1 \\
      & street typology & narrow, commercial, residential streets & 42 & 4.2 & 74 & 6.3 \\
      & \sh transit stations & \sh train stations, bus stops & \sh 32 & \sh 3.2 & \sh 28 & \sh 2.4 \\
      & parking lots & --- & 31 & 3.1 & 41 & 3.5 \\
      & \sh construction sites & \sh including empty plots of land & \sh 14 & \sh 1.4 & \sh 25 & \sh 2.1 \\
      & parks & --- & 6 & 0.6 & 3 & 0.3 \\
      & \sh paving & \sh tiled streets, stone paving & \sh 3 & \sh 0.3 & \sh 9 & \sh 0.8 \\
      & misc & highway, tunnel, pedestrian bridge, train tracks, visible from afar & 4 & 0.4 & 5 & 0.4 \\
    \midrule
    \multirow{4}{=}{Signage}
      & \sh signage (wayfinding) & \sh street names, station names & \sh 59 & \sh 5.8 & \sh 21 & \sh 1.8 \\
      & signage (store) & detached store signs, menus, pricing & 21 & 2.1 & 23 & 2.0 \\
      & \sh signage (advertisement) & \sh campaign posters, festival posters & \sh 8 & \sh 0.8 & \sh 12 & \sh 1.0 \\
      & signage (other) & generic signs, security camera, no-smoking & 7 & 0.7 & 31 & 2.6 \\
    \midrule
    \multirow{4}{=}{Street Objects}
      & \sh trees/plants & \sh trees, non-potted plants & \sh 17 & \sh 1.7 & \sh 25 & \sh 2.1 \\
      & vending machines & --- & 11 & 1.1 & 11 & 0.9 \\
      & \sh street furniture & \sh bulletin board, post box, traffic cone & \sh 9 & \sh 0.9 & \sh 7 & \sh 0.6 \\
      & street art & graffiti & 2 & 0.2 & 7 & 0.6 \\
    \midrule
    \multirow{3}{=}{Transient Objects}
      & \sh parked cars & \sh --- & \sh 15 & \sh 1.5 & \sh 10 & \sh 0.9 \\
      & parked bikes & --- & 3 & 0.3 & 10 & 0.9 \\
      & \sh pedestrians & \sh including pets & \sh 1 & \sh 0.1 & \sh 22 & \sh 1.9 \\
    \midrule
    \textbf{Total} & & & \textbf{1009} & & \textbf{1170} & \\
  \end{tabular}
  \vspace{1em}
  \caption{Distribution of visual elements with example descriptions. Percentages indicate the proportion within each group.}
  \Description{Coding table of visual elements grouped into five categories, each listing tags with example descriptions and the raw count and percentage of mentions in the good navigator group and the poor navigator group.}
  \label{tab:qual-coding-results}
\end{table*}

\clearpage

\section{VLM Pipeline Prompts}

\subsection{Detection Prompt}
\label{sec:detection-prompt}
\begin{promptbox}
You are a visual analyst for a VR navigation tool designed to help people with poor
sense of direction.

You will receive a sequence of $NUM_IMAGES street-level images, taken at intervals
while walking along a city street. The images are named image1 through image$NUM_IMAGES
and are provided in walking order (image1 is earliest, image$NUM_IMAGES is latest).

Each image shows a forward-facing view centered on the walking direction. The images
may be cropped from wider panoramas, so the left and right edges represent the
boundary of the visible area, not the full scene.

Your task is to identify and locate permanent, visually distinctive objects and
landmarks that would help a person orient themselves and remember their route.

--------------------------------------------------

OBJECT CATEGORIES (ranked by baseline priority, highest first)

IMPORTANT - Relative distinctiveness:
The ranking reflects general importance, but actual priority depends on context.
A lower-ranked object that is the ONLY distinctive feature in a scene is more
valuable than a higher-ranked object surrounded by many strong cues.

1. Famous or obvious landmarks - named buildings, monuments, towers, sculptures,
   temples, shrines, bridges, or any structure a local would recognize by name.
2. Wayfinding signage - street names, maps, directional signs to districts or stations.
   Do NOT include generic traffic signs.
3. Store signage (permanent) - brand logos and signage attached to buildings.
   Do NOT include temporary signs or banners.
4. Convenience stores (konbini) - 7-Eleven, FamilyMart, Lawson, etc.
5. Parks and greenspaces - park entrances, gardens.
6. Transit stations and infrastructure - subway entrances, bus stops, station facades.
7. Street furniture - distinctive benches, fountains, clocks, mailboxes.
8. Vending machines - especially clusters or prominent placements.
9. Distinctive non-residential buildings - unusual architecture or strong colors.
10. Other urban infrastructure - overpasses, tunnels, ramps, retaining walls.

--------------------------------------------------

OBJECTS TO EXCLUDE

- Transient objects: cars, bikes, pedestrians, animals.
- Generic regulatory signage: stop signs, speed limits, traffic lights.
- Safety and utility signage: fire hydrant signs, AED signs, pole labels.
- Public-order signs on poles: no-smoking, anti-littering, etc.
- Temporary signage: ads, political posters, event banners.
- Temporary commercial signage not attached to buildings.
- Decorative micro-features: small details only visible up close.
- Any object likely to be moved or removed.

--------------------------------------------------

DETECTION RULES

Rule 1 - Per-frame cap:
- At most 2 objects per image.
- If $NUM_IMAGES = 1 (intersection), you may highlight up to 3 objects.

Rule 2 - Directional bias:
- Prefer objects near the center (walking direction).
- Only select edge objects if:
  * They are major landmarks, OR
  * Highly distinctive, OR
  * The only good candidates.

--------------------------------------------------

SCORING AND CLASSIFICATION

For each object, provide:

category (1-10): category number
category_label: one of:
  "famous landmark", "wayfinding signage", "store signage", "konbini",
  "park/greenspace", "transit station", "street furniture",
  "vending machine", "distinctive building", "urban infrastructure"

color_prominence (0.0-1.0):
- How much the color stands out from surroundings

shape_prominence (0.0-1.0):
- How much the shape stands out from surroundings

semantic_familiarity (0.0-1.0):
- How easy it is to name or describe

All scores should be rounded to 2 decimal places.

--------------------------------------------------

BOUNDING BOX FORMAT

Use: [y_min, x_min, y_max, x_max] in range 0-1000.

IMPORTANT - Box the MOST DISTINCTIVE FEATURE, not the entire object.

Examples:
- Building with mural -> box the mural
- Shrine -> box the gate
- Store -> box the logo/sign
- Station -> box the sign

Only box the whole object if the whole object is the distinctive feature.

--------------------------------------------------

OUTPUT FORMAT

Return ONLY valid JSON with this structure:

{
  "image1": [
    {
      "bbox": [y_min, x_min, y_max, x_max],
      "description": "Describe the object and why it is useful for navigation.",
      "category": 4,
      "category_label": "konbini",
      "color_prominence": 0.75,
      "shape_prominence": 0.40,
      "semantic_familiarity": 0.95,
      "permanence": 0.80,
      "navigational_context": "Explain its role in navigation."
    }
  ],
  "image2": [],
  "image3": []
}

Notes:
- Every image key must be present.
- Each image has 0-2 detections (or up to 3 if single image).
- The same real-world object can appear at most twice across all images.
- description: 1-2 sentences (what + why useful)
- navigational_context: 1-2 sentences (where + role in route)
\end{promptbox}

\subsection{Validation Prompt}
\label{sec:validation-prompt}
\begin{promptbox}
You are a visual analyst validating object detections from a VR navigation tool.

You will receive a single cropped image from a 360-degree panoramic street photo.
A translucent green rectangle is drawn on the image to indicate where the object was detected.
The surrounding area provides additional context.

A previous detection pass produced the following for this region:

- Category: $CATEGORY
- Description: $DESCRIPTION

Your task is to use all three signals - the category, the description, and the
bounding box location - together to determine whether the detection is valid.
The described object does not need to be strictly inside the green rectangle;
it is fine if the object is nearby or partially outside the box.
The bounding box is an approximate guide, not a precise boundary.

--------------------------------------------------

VALIDATION CRITERIA

The detection is VALID if ALL of the following are true:

1. The described object is visible in or near the crop.
   The object mentioned in the description must be identifiable somewhere in the image.
   It does not need to be entirely within the green rectangle - partial overlap or proximity is fine.

2. The category, description, and bounding box point to the same real-world object.
   Minor inaccuracies in the description are acceptable as long as the object is recognizable.
   Reject only for a fundamental mismatch (e.g., "red torii gate" but the image shows a vending machine).

3. The object is permanent.
   It should be a fixed, long-term feature of the environment - not a temporary sign,
   parked vehicle, pedestrian, seasonal decoration, or movable item.

4. The object is visually distinctive enough to serve as a navigation landmark.
   It should be something a person could recognize and describe.
   Generic objects (e.g., a plain concrete wall, a generic utility pole) are not valid.

--------------------------------------------------

The detection is INVALID if ANY of the following are true:

- The crop is empty, blurry, or too dark to identify any object.
- The described object is not visible anywhere in the image.
- The category, description, and bounding box are fundamentally contradictory.
- The object is transient (vehicle, person, temporary sign, etc.).
- The object is too generic or unremarkable to serve as a navigation aid.
- The object is generic street furniture (e.g., street light poles, traffic lights,
  utility poles, guardrails, bollards, metal fences, manhole covers, drainage grates,
  convex traffic mirrors, or similar infrastructure).
- The object is a generic regulatory or public-order sign (e.g., stop signs, yield signs,
  traffic signs, no-smoking signs, pedestrian signs).
- The object is a temporary banner or sign attached to a pole (e.g., event banners,
  promotional flags, shopping district pennants).
- The crop captures only a small, unrecognizable fragment of the object.
- The detection highlights a generic building facade with no distinctive feature.

--------------------------------------------------

OUTPUT FORMAT

Return ONLY a valid JSON object with no additional text:

{
  "is_valid": true or false,
  "reason_is_valid": "One or two sentences explaining your judgment. If valid, briefly confirm what you see. If invalid, explain what is wrong."
}
\end{promptbox}

\section{User Evaluation Materials \& Additional Analysis}

\subsection{Post-Route Questionnaire}

Participants rated the following statements on a 7-point Likert scale (1 = Strongly Disagree, 7 = Strongly Agree):

\begin{enumerate}
    \item I am confident in my scene recognition responses.
    \item I am confident in my intersection recall responses.
    \item I am confident that my mental map accurately represents the route.
    \item I feel I could give directions to someone visiting this area for the first time.
    \item I felt aware of my surroundings during the route.
    \item I feel I could find my way back to the starting point without turn-by-turn directions.
    \item Navigating the route required a significant amount of mental effort.
\end{enumerate}

\noindent Participants also rated their prior familiarity with the route on a scale from 1 to 10, where 1 indicated \textit{``completely new to me; I have never walked here before''} and 10 indicated \textit{``I know this route very well; I walk here regularly.''}

\subsection{Post-Study Questionnaire}
Participants were first asked: \textit{``Which route would you remember better if you had to navigate it again tomorrow?''} and selected either Route~A or Route~B.

\noindent They then rated the following statements on a 7-point Likert scale (1 = Strongly Disagree, 7 = Strongly Agree):

\begin{enumerate}
    \item The circle highlights were easy to notice.
    \item The circle highlights were distracting.
    \item The arrows helped me find highlights I had not yet noticed.
    \item Dismissing the circles by looking at them felt natural.
    \item The highlighting system improved my navigation experience.
    \item I learned how to interact with the highlighting system very quickly.
    \item I would use such a highlighting system with AR glasses in real life.
\end{enumerate}

\newpage
\subsection{Participant Table}
\begin{table*}[h]
  \centering
  \small
  \setlength{\tabcolsep}{5pt}
  \setlength{\aboverulesep}{0pt}
  \setlength{\belowrulesep}{0pt}
  \setlength{\extrarowheight}{2.5pt}
  \renewcommand{\arraystretch}{1.15}
  \rowcolors{2}{white}{rowgray}
  \begin{tabular}{@{}M{0.045\linewidth} M{0.042\linewidth} M{0.075\linewidth} M{0.08\linewidth} M{0.195\linewidth} M{0.13\linewidth} M{0.145\linewidth} M{0.145\linewidth}@{}}
    \textbf{PID} & \textbf{Age} & \textbf{Gender} & \textbf{SBSOD} & \textbf{Languages Fluent In} & \textbf{Japanese Reading} & \textbf{Route 1 Familiarity} & \textbf{Route 2 Familiarity} \\
    \toprule
    E1 & 30 & F & 2.8 & Chinese, English, Japanese & Quite well & Never been there & Never been there \\
    E2 & 28 & M & 2.3 & Chinese, English           & Moderately & Never been there & Never been there \\
    E3 & 25 & F & 2.9 & Chinese, English           & Quite well & Never been there & Never been there \\
    E4 & 31 & M & 2.3 & Chinese, English           & Moderately & Never been there & Never been there \\
    E5 & 28 & F & 3.4 & English, Chinese           & Moderately & Never been there & Never been there \\
    E6 & 25 & F & 3.1 & Chinese, Japanese, English & Very well  & Never been there & Never been there \\
    E7 & 27 & M & 2.9 & Japanese, English          & Very well  & Never been there & Never been there \\
    E8 & 25 & F & 2.3 & English                    & Slightly   & Never been there & Never been there \\
  \end{tabular}
  \caption{Participant demographics, SBSOD scores, and self-reported familiarity with the two study routes.}
  \Description{Table of eight study participants listing age, gender, SBSOD score, languages spoken fluently, self-rated Japanese reading ability, and prior familiarity with each of the two study routes.}
  \label{tab:evaluation-participants}
\end{table*}

\begin{figure*}[ht]
    \centering
    \includegraphics[width=0.95\linewidth]{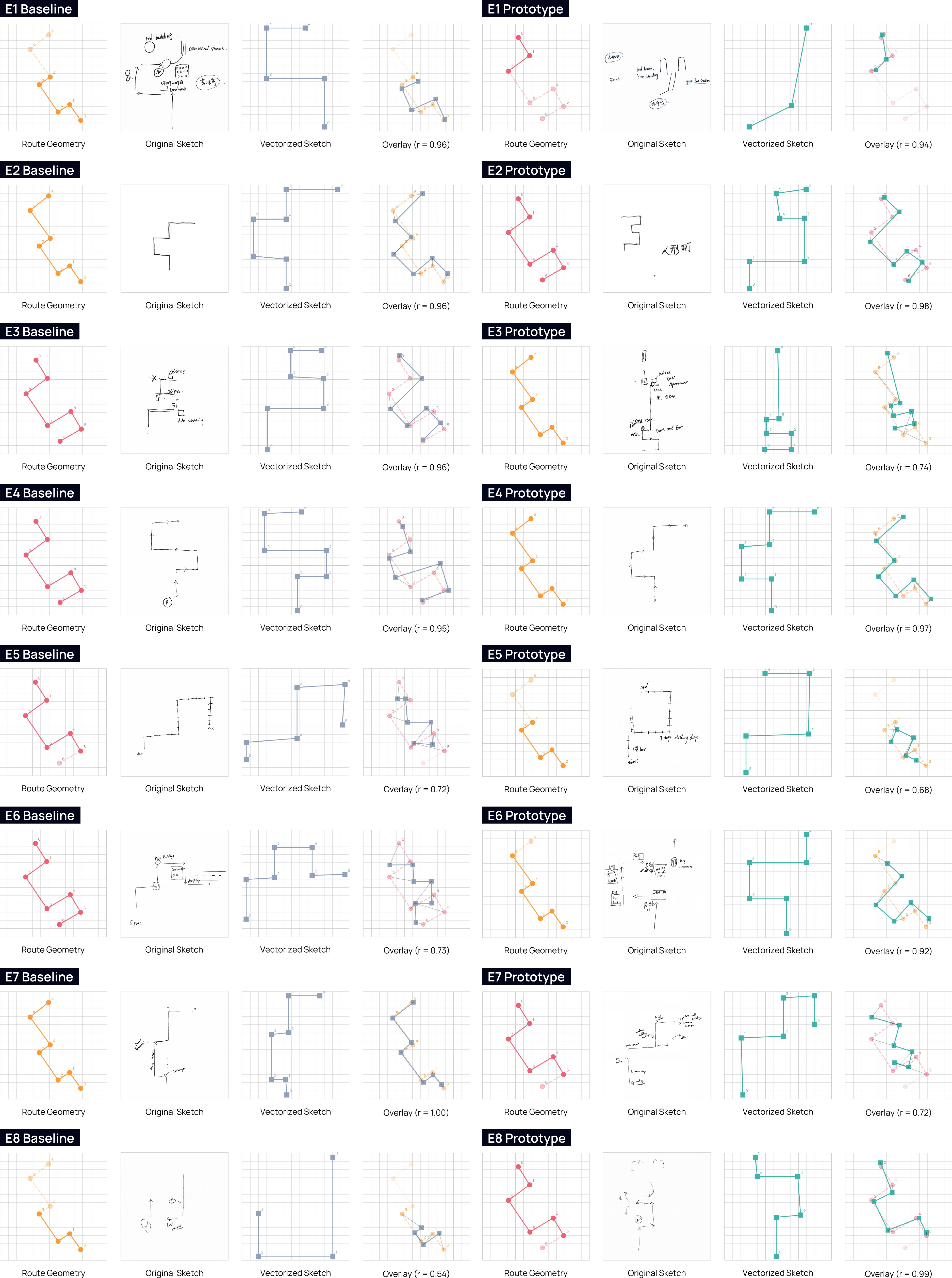}
    \caption{Map sketching task results. 
    For each condition, columns show: the actual route geometry, the participant's original hand-drawn sketch, the vectorized representation of the sketch, and the overlay of the vectorized sketch on the route geometry with the bidimensional correlation coefficient ($r$). Higher $r$ values indicate greater spatial correspondence between the sketched and actual route geometries.
    }
    \Description{Eight-row grid comparing Baseline and Prototype map sketching results for participants E1 to E8. Each condition shows four panels: the actual route geometry, the participant's hand-drawn sketch, its vectorized form, and an overlay of the two with a correlation coefficient. Baseline r values range from 0.54 to 1.00; Prototype r values range from 0.68 to 0.99.}
    \label{fig:evaluation-sketch}
\end{figure*}

\end{document}